\documentclass[pra,superscriptaddress,twocolumn]{revtex4}
\usepackage{amsthm}
\usepackage{amsmath}
\usepackage{mathrsfs}
\usepackage{amssymb}
\usepackage{wasysym}
\usepackage{graphicx}
\usepackage{varwidth}
\usepackage{url}
\usepackage{dsfont}
\usepackage{float}
\usepackage{bm}
\usepackage{ifthen}
\usepackage[usenames,dvipsnames]{color}
\usepackage{mathrsfs}
\usepackage[colorlinks=true,citecolor=blue,urlcolor=black]{hyperref}
\usepackage{float}
\usepackage{afterpage}
\usepackage{subfigure}

 \usepackage{caption}
\usepackage{algorithm, algorithmic}%
\usepackage{xcolor}
 \newcommand{\fp}{\ensuremath{f_{\text{pl}}}}
 \newcommand{\ftls}{\ensuremath{ \omega_{\text{TLS}}}}
 \newcommand{\tr}{\ensuremath{t_{\text{r}}}}

 \newcommand{\tp}{\ensuremath{t_{\text{p}}}}
 
\newcommand{\sket}[1]{{\ensuremath{\lvert#1\rangle}}}
\newcommand{\lket}[1]{{\ensuremath{\left\lvert#1\right\rangle}}} 
\newcommand{\ket}[1]{\if@display\lket{#1}\else\sket{#1}\fi}
\newcommand{\sbra}[1]{{\ensuremath{\langle#1\rvert}}}
\newcommand{\lbra}[1]{{\ensuremath{\left\langle#1\right\rvert}}}
\newcommand{\bra}[1]{\if@display\lbra{#1}\else\sbra{#1}\fi}
\newcommand{\sbraket}[2]{{\ensuremath{\langle#1\rvert#2\rangle}}}
\newcommand{\lbraket}[2]{{\ensuremath{\left\langle#1\!\left\rvert\vphantom{#1}#2\right.\!\right\rangle}}}
\newcommand{\braket}[2]{\if@display\lbraket{#1}{#2}\else\sbraket{#1}{#2}\fi}

\newcommand{\sketbra}[2]{{\ensuremath{\lvert #1\rangle\!\langle #2\rvert}}}
\newcommand{\lketbra}[2]{{\ensuremath{\left\lvert #1\right\rangle\!\!\left\langle #2\right\rvert}}}
\newcommand{\ketbra}[2]{\if@display\lketbra{#1}{#2}\else\sketbra{#1}{#2}\fi}

\theoremstyle{plain}

\theoremstyle{definition}

\begin{document}
\title{ Learning Non-Markovian Quantum Noise from  Moir\'{e}-Enhanced Swap Spectroscopy with Deep Evolutionary Algorithm}
\author{Murphy Yuezhen Niu, Vadim Smelyanskyi}
\affiliation{Google Research, 340 Main Street, Venice Beach, California 90291, USA}
\author{  Paul Klimov}
\affiliation{Google Research, 6868 Cortona Dr, Goleta, California 93117, USA}
\author{Sergio Boixo}
\affiliation{Google Research, 340 Main Street, Venice Beach, California 90291, USA}
\author{Rami Barends,   Julian Kelly, Yu Chen,    Kunal Arya,  
  Brian Burkett, Dave Bacon, Zijun Chen}
  \affiliation{Google Research, 6868 Cortona Dr, Goleta, California 93117, USA}
\author{Ben Chiaro}
\affiliation{Department of Physics, University of California, Santa Barbara, California 93106-9530 USA}
\author{
Roberto Collins,  Andrew Dunsworth,  Brooks Foxen,  Austin Fowler, Craig Gidney, Marissa Giustina, Rob Graff,      Trent Huang,  Evan Jeffrey,    David Landhuis,   Erik Lucero,   Anthony Megrant, Josh Mutus, Xiao Mi, Ofer Naaman, Matthew Neeley, Charles Neill,  Chris Quintana, Pedram Roushan}
\affiliation{Google Research, 6868 Cortona Dr, Goleta, California 93117, USA}

\author{John M. Martinis}
\affiliation{Google Research, 6868 Cortona Dr, Goleta, California 93117, USA}

\affiliation{Department of Physics, University of California, Santa Barbara, California 93106-9530 USA}
\author{Hartmut Neven}
\affiliation{Google Research, 340 Main Street, Venice Beach, California 90291, USA}


%
\begin{abstract}
Two-level-system~(TLS) defects in amorphous dielectrics are a major source of noise and decoherence in solid-state qubits. Gate-dependent non-Markovian errors caused by TLS-qubit coupling are detrimental to fault-tolerant quantum computation and have not been rigorously treated in the existing literature. In this work, we derive the non-Markovian dynamics between TLS and qubits  during a SWAP-like two-qubit gate  and the associated average gate fidelity for frequency-tunable Transmon qubits.  This gate-dependent error model facilitates using qubits  as sensors to simultaneously learn  practical imperfections in both the qubit's environment and control waveforms. We combine the-state-of-art machine learning algorithm with Moir\'{e} enhanced swap spectroscopy to achieve robust learning using noisy experimental data. Deep neural networks are used to represent  the functional map from  experimental data to  TLS  parameters, and are trained through an evolutionary algorithm. Our method achieves the highest  learning efficiency and robustness against experimental imperfections to-date, representing an important step  towards in-situ quantum control optimization over environmental and control defects. 

\end{abstract}
  
\maketitle

%
%
%




Two-level-system~(TLS) defects in amorphous dielectrics  are   a major source of noise and decoherence in superconducting qubits~\cite{muller2017towards}.  Substantial progress has been made towards understanding TLS microscopic origin~\cite{Yu2004,Cole2010,Leggett2013},  statistical properties~\cite{Martinis2005,Shalibo2010,Barends2013}, and   their mutual interactions~\cite{Lisenfel2015,Klimov2018}   using   superconducting qubits as probes. In particular, the  physical properties of TLS manifest in their interaction dynamics with coupled qubits, which can be measured when the qubit frequency is near resonant with the TLS.
In frequency tunable qubits, multi-qubit gates are executed by sweeping the participating qubits to near resonance~\cite{barends2014superconducting}.  Over the course of the two-qubit-gate frequency trajectories, the participating qubits are susceptible to TLS induced gate errors, the most significant of which are non-Markovian. However,  there lacks physical models for the non-Markovian errors  induced by qubit-TLS interaction during quantum gates. Consequently, little is known about a truthful  noisy quantum channel description for existing quantum gates under the influence of TLSs. Such a description, however, is essential for developing  real-time  error characterization and error mitigation schemes to combat TLS fluctuations, which can happen on timescales ranging from minutes to hours~\cite{Klimov2018}.


In this work, we develop an experimentally relevant model of non-Markovian   dynamics  of charge-charge  TLS-qubit interactions, with weakly coupled Markovian environments for TLS and qubits. This model is then deployed to an efficient TLS characterization method  for  frequency-tunable superconducting Transmon qubits using noisy  swap-spectroscopy data~\cite{Klimov2018}. We utilize a  deep-neural-network based evolutionary algorithm~(DNN-EA) for the robust  learning of TLS   parameters. The efficiency of our characterization is further improved by harnessing the Moir\'{e} effect through non-uniform temporal sampling when collecting the swap-spectroscopy data~\cite{Miao2016}. The Moir\'{e} effect  amplifies the periodicity of the non-Markovian interference pattern and reduces the number of data points and thus overall runtime necessary for the TLS model learning. We  achieve a two-order-of-magnitude higher accuracy of experimental characterization against measurement noise while reducing the required optimization time  by a factor of $10^4$ over exhaustive search. Lastly, we provide an  operator sum description and the associated average  fidelity of a noisy two-qubit gate under an experimentally observed TLS-qubit interaction. Our model predicts a significant degradation in average two-qubit gate fidelity   when the TLS frequency is in the viscinity of the two-qubit gate interaction frequency.
 The gate-dependent error model developed and characterized in this work points to a new direction of using qubits  as sensors to   simultaneously  characterize  both the practical imperfections in the qubit environment and the  non-idealities in qubit control waveforms. This lays the foundation for developing next generation   quantum devices with improved fidelity and robustness against various experimental imperfections.

\begin{figure}[H]
\begin{center}
\includegraphics[width=0.7\linewidth]{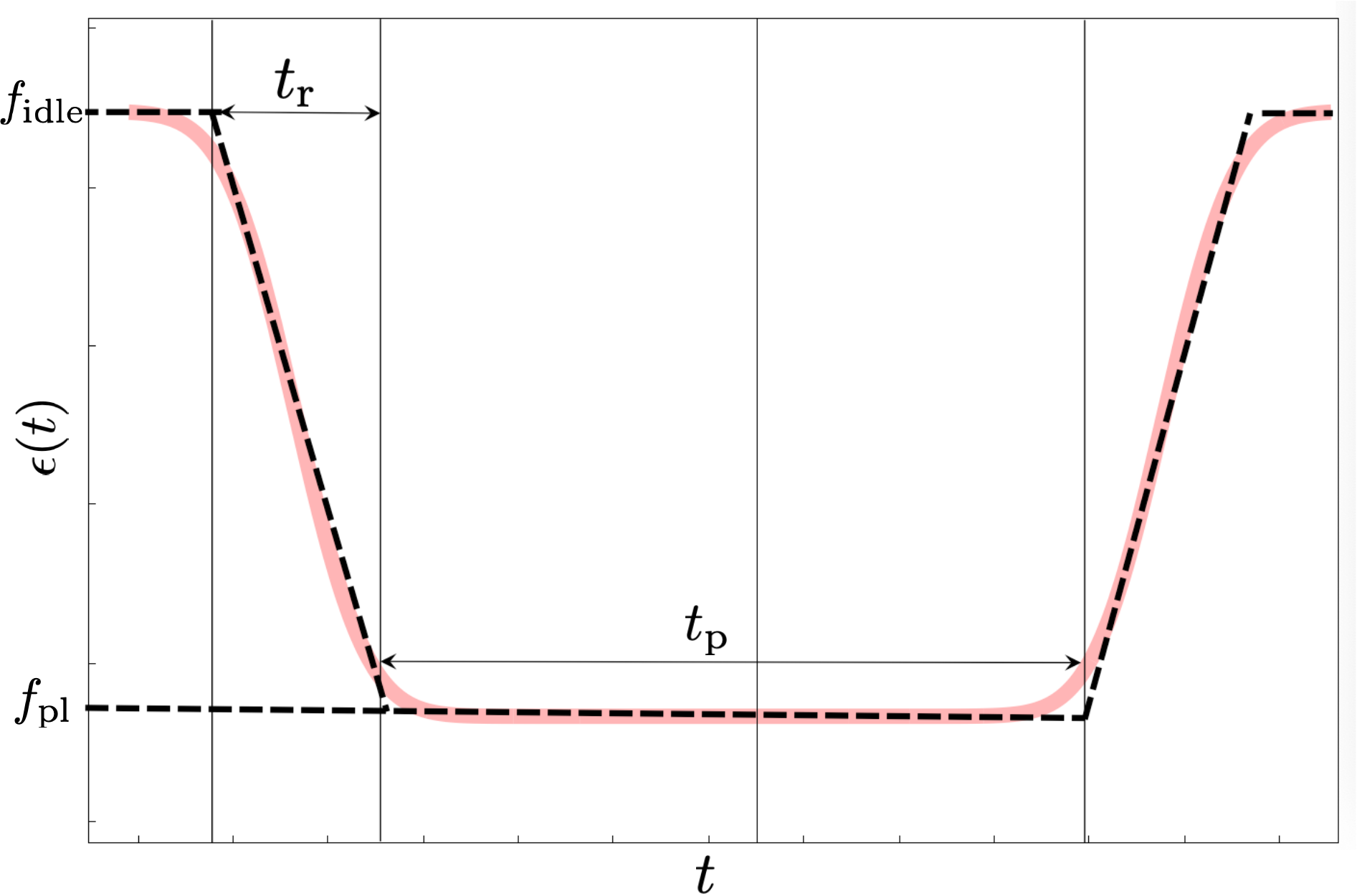}
\caption{The qubit frequency control modulation $\epsilon(t)$ as a function of time $t$  during swap spectroscopy.  
\label{PulseShapeSwapspec}}
\end{center}
\end{figure}

Swap spectroscopy is a powerful tool for measuring  the spectral and temporal properties of individual TLS  coupled to qubits~\cite{Barends2013,Klimov2018}.  During the swap spectroscopy measurement,  qubit evolves under the time-dependent control Hamiltonian $\hat{H}_{Q,0}(t)=-\epsilon(t)/2\sigma^z$, where $\sigma^z$ represents  Pauli Z operator of  qubit. The shape of this frequency modulation $\epsilon(t)$ resembles a smoothed tapezoid, see Fig.~\ref{PulseShapeSwapspec}， which is parametrized by initial  frequency $f_{\text{idle}}$,   plateau frequency  $f_{\text{pl}}$,   ramp time \tr, and hold time \tp. 
We name the corresponding unitary transformation a swap-spectroscopy gate: $U_{\text{swap}}$. 
The two-dimensional swap spectroscopy data~(TSSD) are probabilities $\{ P_{\text{decay}}^{\text{exp}}(t_{\text{p}},    f_{\text{pl}})\}$ of  a qubit decaying to its ground state from an initial excited state after the application of  $U_{\text{swap}}$  for a range of plateau frequencies and wait times  $\{f_{\text{pl}}, t_{\text{p}}\}$, where  $t_{\text{r}}$ is usually fixed and determined by the frequency bandwidth of  control electronics.

\begin{figure}[H]
\begin{center}
\includegraphics[width=1\linewidth]{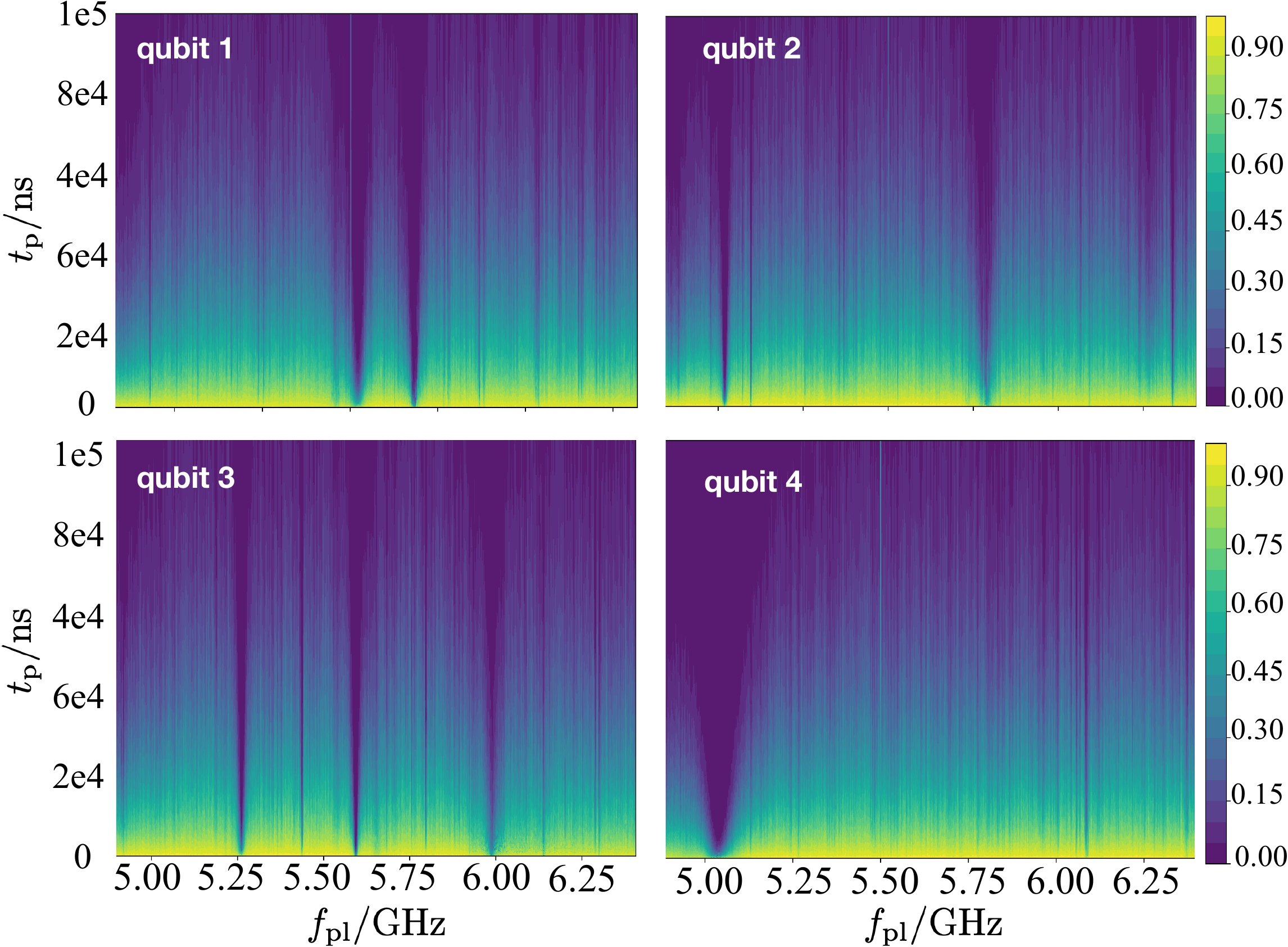}
\caption{Experimental measurements of $\{1-P_{\text{decay}}^{\text{exp}}(\tp, f_{\text{pl}})\}$ for four  qubits from the same quantum processor: probabilities of remaining in  the excited state after a swap-spectroscopy gate as a function of    frequency  and hold time.  Accelerated decay caused by the coupling of qubits to environmental defects manifest as deep blue lines. 
\label{fourqubits}}
\end{center}
\end{figure}

TSSD  bares the subtle footprints of the underlying incoherent and coherent errors of a quantum gate. This is because  the  frequency control for Google's frequency-tunable qubits during  a two-qubit gate~\cite{Martinis2010}  takes similar form as the swap-spectroscopy gate in Fig.~\ref{PulseShapeSwapspec}. From experimental observations, we learn that such TLS-induced noise is prevalent across the frequency modulation range of existing qubits~\cite{Klimov2018}, see for example Fig.~\ref{fourqubits} for data obtained on  four frequency tunable qubits from the same quantum chip: each deep vertical blue line extending into the yellow region~(where qubit state  is largely unscathed) manifests an unwanted population transfer between each qubit and its  environment.

One key ingredient has been missing to directly utilize TSSD for  noisy quantum channel characterization:  a  physical model for the qubit and TLS interaction dynamics as a function of quantum gate parameters.  We derive an experimentally relevant  non-Markovian model of  TLS-qubit dynamics, which encompasses two limiting cases: 1. The TLS and qubit are weakly coupled such that   the Born approximation and the second order linear perturbation theory apply,  and  the coupling between the TLS and it's environment can  be treated as Markovian; 2. The TLS and qubit are strongly coupled and the joint system is largely coherent during the evolution of a quantum gate.





In our  model, the  Hamiltonian of the joint TLS-qubit-bath system comprises three parts:
\begin{align}
&\hat{H}(t) =  \hat{H}_{TLS}(t) + \hat{H}_Q(t) + \hat{H}_{Q-TLS}(t) 
\end{align}
where we use $ \hat{H}_{TLS}(t) $ to represent TLS's free Hamiltonian, its coupling to an environment and the environmental Hamiltonian; similarly    $\hat{H}_Q(t)$ includes both control Hamiltonian $\hat{H}_{Q,0}(t)$ and qubit's coupling to environmental defects~(see details  in Supp.~\ref{SecLinearPerturbation}).  Swap spectroscopy  measures qubit relaxation induced by  transversal charge-charge coupling between the qubit and TLS of the form:  $\hat{H}_{TLS-Q} = \lambda  \sigma^x_{TLS}\sigma_Q^x $, where we use $\lambda$ as TLS-qubit coupling strength and $\sigma_i^k$ as the Pauli $k$ operator of $i$'s system.

We first consider the weak coupling limit where: 1. the TLS-qubit coupling is much smaller than the inverse of the quantum gate time  $\lambda \ll 2\pi/t_{\text{p}}$ such that linear perturbation theory applies; 2. $\lambda$ is sufficiently large compared to the TLS decay rate which violates  the Markovian constraints. We   solve the dynamics of the joint TLS-qubit  system using only Born  approximation~(joint system of qubit and TLS remains in a product state) while abandoning the commonly adopted Markov approximation~(dynamics is memoriless).  Consequently, the system dynamics can no longer be represented by master equations of Lindbladians.
 
The initial state of TLS-qubit system is taken to be a product state  $  \rho(0)=\ket{1}\bra{1}_Q \otimes \ket{0}\bra{0}_{TLS}$ based on the fact that TLS is most likely to be in its ground state due to the much shorter coherence time  than that of a qubit. Moving into the interaction picture $\tilde{\rho}(t)=U_Q^\dagger \rho(t)U_Q $ defined by the frame rotation $U_Q =\mathcal{T}[\int_0^t e^{-i \hat{H}_Q(\tau)}d\tau]$.  Tracing out the TLS subsystem gives us the time-dependent qubit density operator~(see Supp.~\ref{SecLinearPerturbation}):
\begin{align}\label{blochRedfield2}
&\tilde{\rho}(t) =\rho_Q(0) - \frac{\lambda^2}{\hbar^2}\int_0^t ds \int_0^s d \tau\left[ C(s-\tau)\tilde{\sigma}_{Q}^x(s)\tilde{\sigma}_{Q}^x(\tau)\rho(0) \right.\\\nonumber
&\left. + C(\tau-s)\tilde{\sigma}_{Q}^x(\tau)\tilde{\sigma}_{Q}^x(s)\rho(0)   
- C(\tau-s) \tilde{\sigma}_{Q}^x(s)\rho(0) \tilde{\sigma}_{Q}^x(\tau)  \right.\\\nonumber
&\left. - C(s-\tau) \tilde{\sigma}_{Q}^x(\tau)\rho(0) \tilde{\sigma}_{Q}^x(s)] \right],
\end{align} 
where the TLS correlator $
C(\tau -s)=  \langle \tilde{\sigma}_{TLS}^x(\tau)  \tilde{\sigma}_{TLS}^x(s)\rangle $   depends on the properties of TLS's environment. Within the   experimental applicability, we choose TLS's response function as that given by  a Markovian coupling between TLS and its environment: $
C(\tau -s) 
=  e^{-\Gamma_{TLS,\phi} |\tau - s| - i \omega_{TLS}(\tau-s)} $, where $\Gamma_{TLS,\phi}$ is the dephasing rate of TLS.
The qubit correlator from Eq.~(\ref{blochRedfield2}) in the interaction picture after tracing out its environment becomes:
\begin{align}\nonumber
\tilde{\sigma}^x_Q(\tau)\tilde{\sigma}^x_Q(s) &=[\cos\phi(\tau)\sigma_Q^x +\sin\phi(\tau)\sigma_Q^y ]\\
&\times[p\cos\phi(s)\sigma_Q^x +\sin\phi(s)\sigma_Q^y] e^{-\Gamma_{2,q} \vert \tau -s \vert } ,  
\end{align}
which depends on both the qubit's dephasing rate $\Gamma_{2,q}$ from the Markovian environment and the dynamical phase accumulated during $U_{\text{swap}}$ as: $
\phi(\tau) =\int_0^\tau \epsilon(t)dt =   f_{\text{idle}} \tau +  \epsilon_m \int_0^\tau \mu(t) dt $ with $\epsilon_m= \fp-f_{\text{idle}}$ and $0\leq \mu(t)\leq 1$, a dimensionless time-dependent function representing the frequency control trajectory of swap-spectroscopy gate~(see Fig.~\ref{PulseShapeSwapspec}).
Inserting these results into the density operator expression gives us the qubit decay  probability $P_{\text{decay}} 
=  \bra{0}_q\tilde{\rho}_q(\tp + 2\tr)\ket{0}_q$ at the end of a swap-spectroscopy gate as:
\begin{align}
&P_{\text{decay}} \nonumber
 = \frac{\lambda^2 t_{\text{p}}^2}{\hbar^2}\text{Re}\left[  \int_0^1 dx\int_0^1dy e^{-t_{\text{p}}\Gamma_{ 2} |x-y|} \right. \\\label{probabilityUnsimplifiedEq}
& \left.\times e^{( i t_{\text{tot}}  \left\{  (x-y)(f_{\text{idle}} - \omega_{TLS})   + \epsilon_m  \int_y^x \mu(z) dz  \right\}} \right]
\end{align}
where  $\Gamma_2=\Gamma_{TLS,\phi} + \Gamma_{2,q}$, and the time is normalized by the overall runtime $ t_{\text{tot}}= 2 \tr + t_{\text{p}}$. Measuring  $P_{\text{decay}}^{\text{exp}}(t_{\text{p}}, f_{\text{pl}})$ as a function of $t_{\text{p}}$ and $f_{\text{pl}}$ with a fixed $\tr$ allows us to reconstruct the TLS model parameters.  Numerically integrating Eq.~(\ref{probabilityUnsimplifiedEq}) is computationally expensive for $10^6$ data points to satisfy different convergence requirements
for both long-time $\sim 1/\Gamma_2$ and short-time $\sim 1/(f_{\text{idle}} - \omega_{TLS})$ dynamics. Instead, we apply the stationary phase approximation to derive a  simplified closed-form expression for Eq.~(\ref{probabilityUnsimplifiedEq}), which takes three orders of magnitude less time to evaluate and depends on six physical parameters: $\lambda, t_{\text{p}}, t_r, \Gamma_2,  f_{\text{idle}}, \omega_{TLS}$, see Supp.~\ref{SecLinearPerturbation}. 

Next, we consider the case  when  $\lambda \gg \Gamma_2$ and   $1/t_{\text{p}}  \gg \Gamma_2$ and the joint system is approximately coherent during the   gate operation. We derive a closed-form expression for the Landau-Zener-Rabi oscillation  under a trapezoidal frequency control pulse, which reproduces the experimentally observed Moir\'{e} pattern~(see Fig.~\ref{CompareMoireUniformFig} and Supp.~\ref{SecCoherentTLSqubit}). 
This coherent interaction model  also elucidates two distinct ways the pulse shape of qubit's frequency control  influences   the qubit-TLS dynamics.  
 
The first route to qubit decay is through the  Landau-Zener~(LZ)    transition during   frequency tuning   parts of the control trajectory, i.e. the ramp up and ramp down portions of Fig.~\ref{PulseShapeSwapspec}.  It occurs    when the  qubit frequency passes the TLS frequency at $\epsilon=0$. The LZ contribution to qubit's decay  
is of order $\delta P_{\text{LZ}}\approx \frac{2\pi \lambda^2}{v} $,  which is around $   10^{-3}$ to $   10^{-5}$ for $g \in [1, 10]$~MHz and $ v\approx 0.1\quad \text{GHz}^2$. 
The second mechanism is the coherent  population oscillations~(Rabi) between  qubit and TLS throughout the whole control trajectory.  The  ramp   portions of the trapezoidal pulse alter the overall phase of this oscillation by roughly $ 2 \lambda (\fp - f_{\text{idle}})/v $ due to the boundary effects at both ends of the plateau, which can be more significant than LZ contribution and non-negligible under certain conditions~(see Supp.~\ref{LZdiscussionSec} for more discussions). 

Notice that  the   LZ effect  is taken under two conditions:  $\sqrt{v}\gg \lambda$  and  $\epsilon_m \gg \lambda$. The second condition guarantees that the boundary effect is negligible and is violated in our  experiment.  The coherent model we developed takes care of both contributions and account  for the complete non-Markovian qubit-TLS dynamics which depends on a larger set of frequency control parameters.  
This increased sensitivity to  the qubit's control pulse shape provides  another intriguing opportunity: qubit can serve as a sensor to   characterize practical non-idealities  from  both the qubit environment and from qubit control actuation simultaneously. 

 



%
%
%
%
%

Learning physical models from swap spectroscopy data, however, is inherently difficult due to the coexistence of quantum dynamics of drastically different time-scales. On the one hand, we have the exponential decay of qubit population due to incoherent coupling with TLSs and a bosonic environment that is  of time scale $\apprge 10\mu$s. To detect such exponential behavior accurately,  swap-spectroscopy data spanning at least 100$\mu$s is needed. On the other hand, the coherent effect of qubit-TLS interaction manifests in a time scale of $10\text{ns} \sim 100\text{ns}$. This is determined by the inverse of frequency gap $\epsilon_m$ between  qubit and TLS and the coupling strength $\lambda$. Due to the Nyquist–Shannon sampling theorem, the time interval between data points should be taken  less than 5ns. For the frequency range~(1GHz) and frequency accuracy~(1MHz) we are interested, this implies a data size of around $10^9$. Since TLS parameters are drifting   in time~\cite{Klimov2018}, online characterization requires time-labeled data which increases the required data size even further. Processing such large amount of data in real time  or between system resets can be impractical. Lastly, experimental imperfections introduce noise to the measured data, which makes the learning of TLS model parameters highly susceptible to errors. 

Facing these major challenges, our goal is to develop a practical TLS characterization method  which meets the following criteria:  (1) avoid underfitting or overfitting, (2) be robust against noise, and (3)   faster than traditional methods  such as exhausted search in identifying the optimal TLS parameters. The last reqiurement is essential to the capability of online characterization of a time-dependent TLS model~\cite{Klimov2018}. 
To achieve these goals   we deploy two important methodologies: spectral amplification   based on Moir\'{e}s effect and modern machine learning algorithms.

To reduce the amount of required experimental data, we harness the  Moir\'{e}'s effect where overlaying two periodic patterns creates a new pattern with a larger period. 
We experimentally collect $P_{\text{decay}}^{\text{exp}}(t_{\text{p}},  f_{\text{pl}})$ for an ensemble of  $f_{\text{pl}}\in $ $[4.6, 5.8]$GHz with a uniform step size  $\delta f_{\text{pl}}=$1MHz, and   $t_{\text{p}}\in[10, 10^5]$ns with a non-uniform logarithmic step size. 
Similar to the effect of Shape Moir\'{e}~\cite{MoireMagnify}, the  classical interference from non-uniform temporal sampling amplifies the \textit{temporal} periodicity of non-Markovian oscillation patterns by three orders of magnitude in swap-spectroscopy data~(see Supp.~\ref{MoireSupp}). Consequently, the   required sampling rate  and the size of the experimental data are reduced commensurately from the Nyquist theorem.

This substantial reduction in the experimental data  size accelerates the process of physical model inference.  To extract the theoretical model parameters,  we  minimize the following cost function:
\begin{align}\nonumber
&\text{C}_{\text{fit}}(\Delta \fp, \Delta \tp, \vec{p}_{\text{TLS}})= \sum_{k\in [\Delta \fp], j\in [\Delta \tp]   }\\\label{meansquareError}
&\vert P_{\text{decay}}^{\text{exp}}(t_{\text{p}}^j,  f_{\text{pl}}^k)  - P_{\text{decay}}(t_{\text{p}}^j, \fp^k,  \vec{p}_{\text{TLS}})  \vert^2 
\end{align}
which is the L2 norm of the difference between   experimental data $ P_{\text{decay}}^{\text{exp}}(t_{\text{p}}^j,  f_{\text{pl}}^k)$ and  predicted values from our physical model for the chosen range of $f_{\text{pl}}\in \Delta \fp$ and $t_{\text{p}}\in \Delta t_{\text{p}}$ given the TLS   parameters $\vec{p}_{\text{TLS}}=\{\Gamma_2, \tr, \lambda, \ftls\}$. Clearly, the less data points there are, the faster it is to evaluate  Eq.~(\ref{meansquareError}). Our closed-form derivation of $P_{\text{decay}}(t_{\text{p}}^j, \fp^k,  \vec{p}_{\text{TLS}})$~(see  Supp.\ref{SecLinearPerturbation}) further shortens the evaluation time of the cost function. 

Two more challenges remain to be addressed: the total data size  of order $10^6$ still greatly exceeds the number of free parameters in the TLS model;   experimental  data contain noise due to measurement errors and various background  fluctuations~(see Supp.~\ref{ExperimentNoiseSec}), which creates false local optimalities in the cost function of Eq.~(\ref{meansquareError}) and makes the learning susceptible to errors. We show that a gradient-free training of deep neural network through evolutionary algorithm tackle both challenges at once.
 
  \begin{figure}[H]
\begin{center}
\includegraphics[width=1\linewidth]{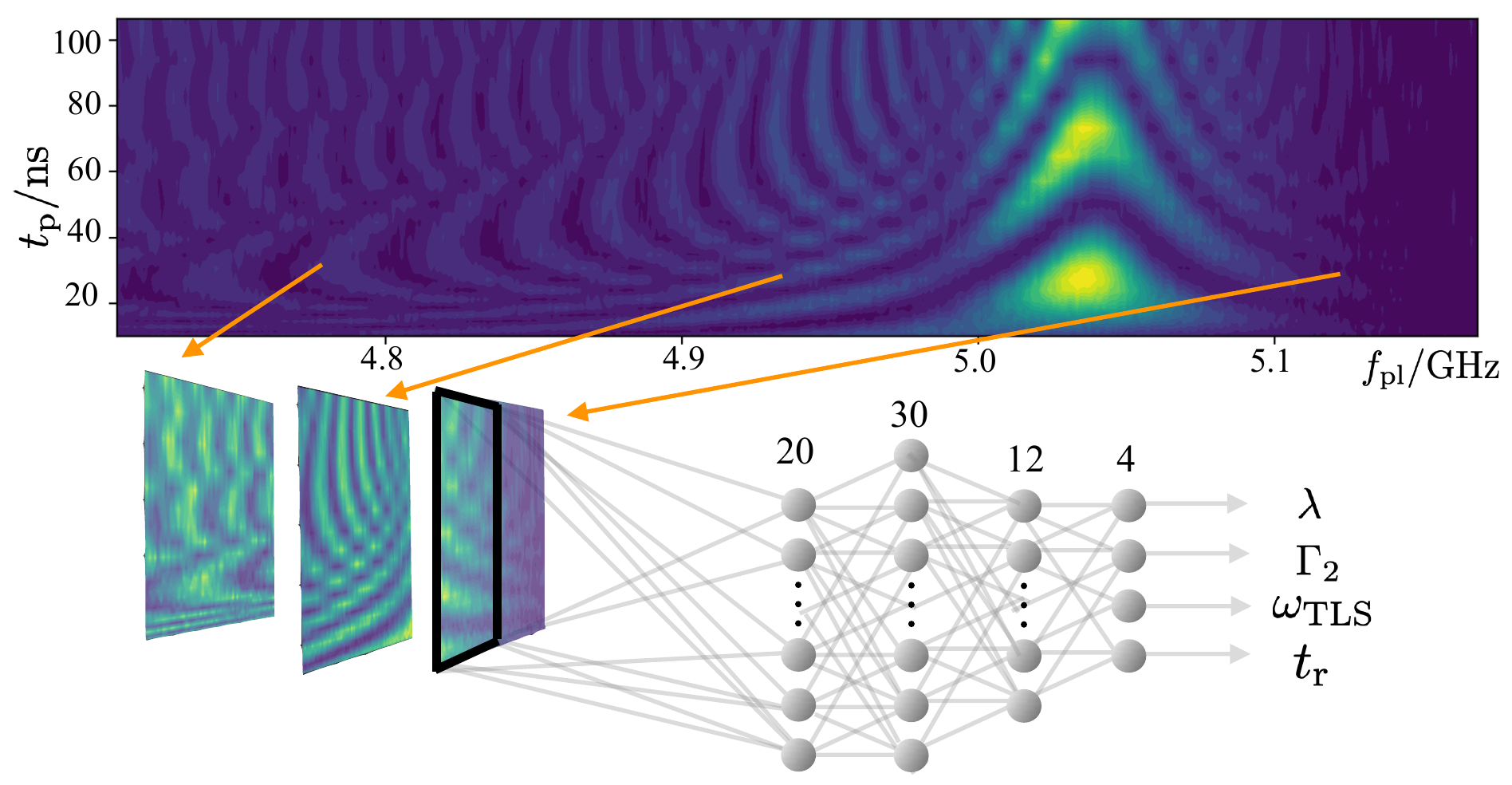}
\caption{Diagram of the DNN learning architecture: each section of the two-dimensional data~(represented by the black frame) from different frequency regions are  input to a four layer fully connected DNN with hidden layer dimensions $20, 30, 12$ and 4. By using an evolutionary algorithm, the last layer of the DNN is trained to output the TLS parameters: $\lambda, \Gamma_2, \ftls,$ and $\tr$ that best reproduce the experimental data given our theoretical model in Eq.~(\ref{probabilityUnsimplifiedEq}).
\label{DNNFIG}}
\end{center}
\end{figure} 

We parametrize the functional relation between TSSD and TLS parameters $\vec{p}_{\text{TLS}}$ using a deep neural network~(DNN), which takes TSSD data as input and outputs $\vec{p}_{\text{TLS}}$. Through such re-parametrization, we can increase the dimensionality of the fitting parameters arbitrarily by choosing a larger  neural network. Detailed implementation is supplied in Fig.~\ref{DNNFIG}  and Supp.~\ref{SuppEA}.
We choose the evolutionary algorithm~(EA) to train the neural network based on its well-known robustness against sample   noise and its obviation  of backpropagation~\cite{salimans2017evolution}. The later helps us to avoid unwanted gradient explosion that is common during DNN training~\cite{pascanu2012understanding}.
EA's efficiency in finding a global optimal solution of TLS parameter is also verified in our numerical optimization.

EA based on DNN~(DNN-EA) can be described by  iterations of the following steps:  1.  initialize the DNN parameters with some random values; 2. perturb the initial guess around a zero mean Gaussian distribution with a chosen variance  to obtain many new DNN configurations; 3. evaluate the cost function  in Eq.~(\ref{meansquareError})  obtained by each newly sampled DNN; 4. update DNN by  averaging over  all the sampled DNNs  weighted by each associated cost. EA is intimately related to conventional sense of reinforcement learning in that it does not need labeled data and learn by iterations of exploration and exploiting these explorations through performance evaluations. Detailed  EA implementation  is prescribed in Supp.~\ref{SuppEA}.

\begin{figure*}[ht]
\begin{center}
\includegraphics[width=1\textwidth]{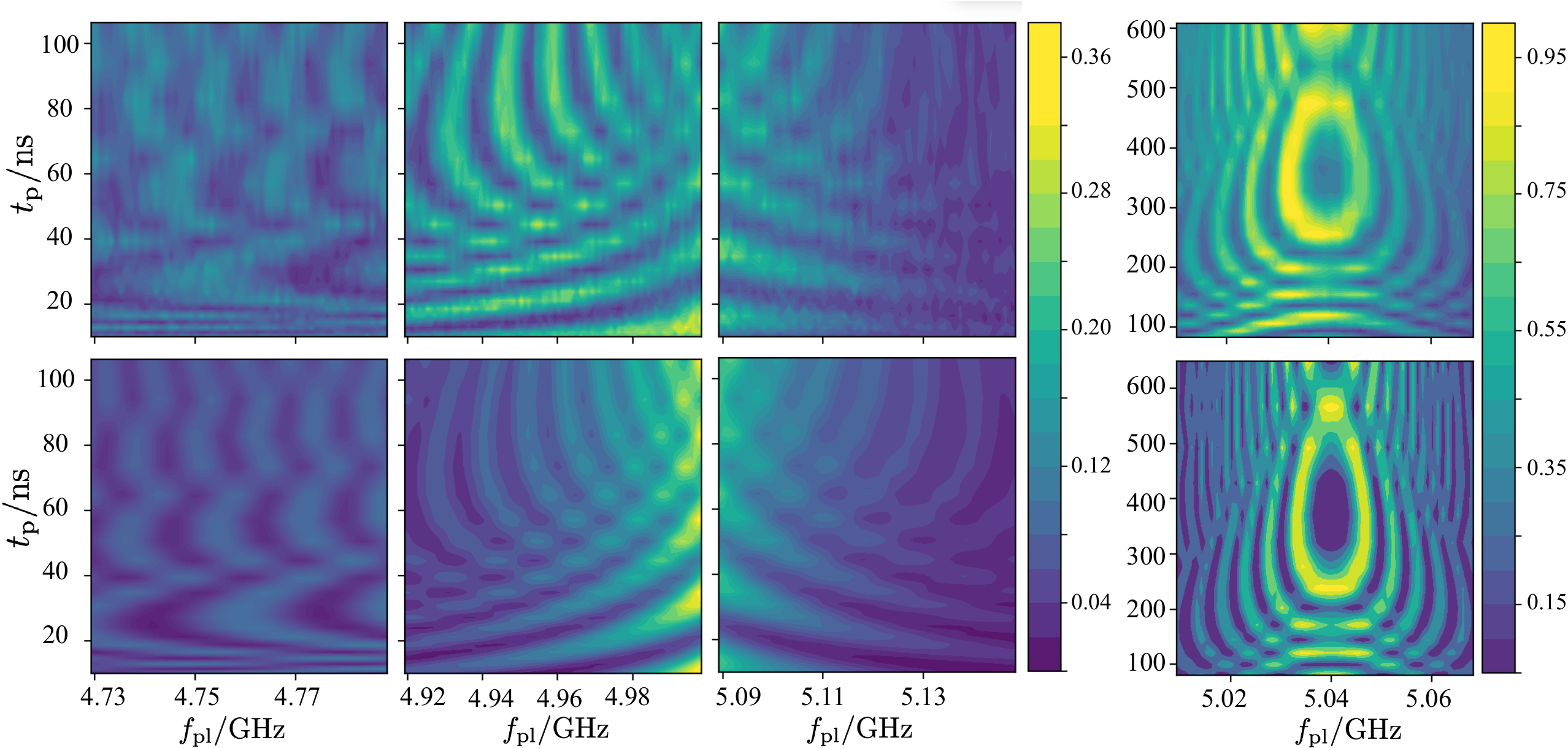}
\caption{Experimentally measured TSSD ~(first row) and  theoretically predicted TSSD from learned model~(second row) with $ f_{\text{idle }}=5.6$ GHz,   $\lambda =8.959$MHz, $\Gamma_2=10$MHz, $\ftls=5.04$GHz, $\tr =6$ns for four different regimes of plateau frequencies:  $\fp \ll \ftls$~(leftmost column), $\fp < \ftls$~(third column from right), $\fp > \ftls$~(second column from right), and $\fp \approx \ftls$~(rightmost column).
\label{sumplotfitexperiment}}
\end{center}
\end{figure*}

We take qubit 4 in Fig.~\ref{fourqubits} as an example~(see Supp.~\ref{SuppData} for the full data), there are four regions of distinct features in TSSD, see Fig.~\ref{sumplotfitexperiment}:  I. $\fp \approx \ftls$, II.  $\fp > \ftls$, III. $\fp < \ftls$, and IV. $\fp \ll \ftls$.
Due to Moir\'{e}'s effect, unlike what has been commonly observed in traditional TLS spectroscopies, our experimental data demonstrates  complex circular interference patterns  in addition to the commonly known Chevron patterns.

\begin{figure}[H]
\begin{center}
\includegraphics[width=1\linewidth]{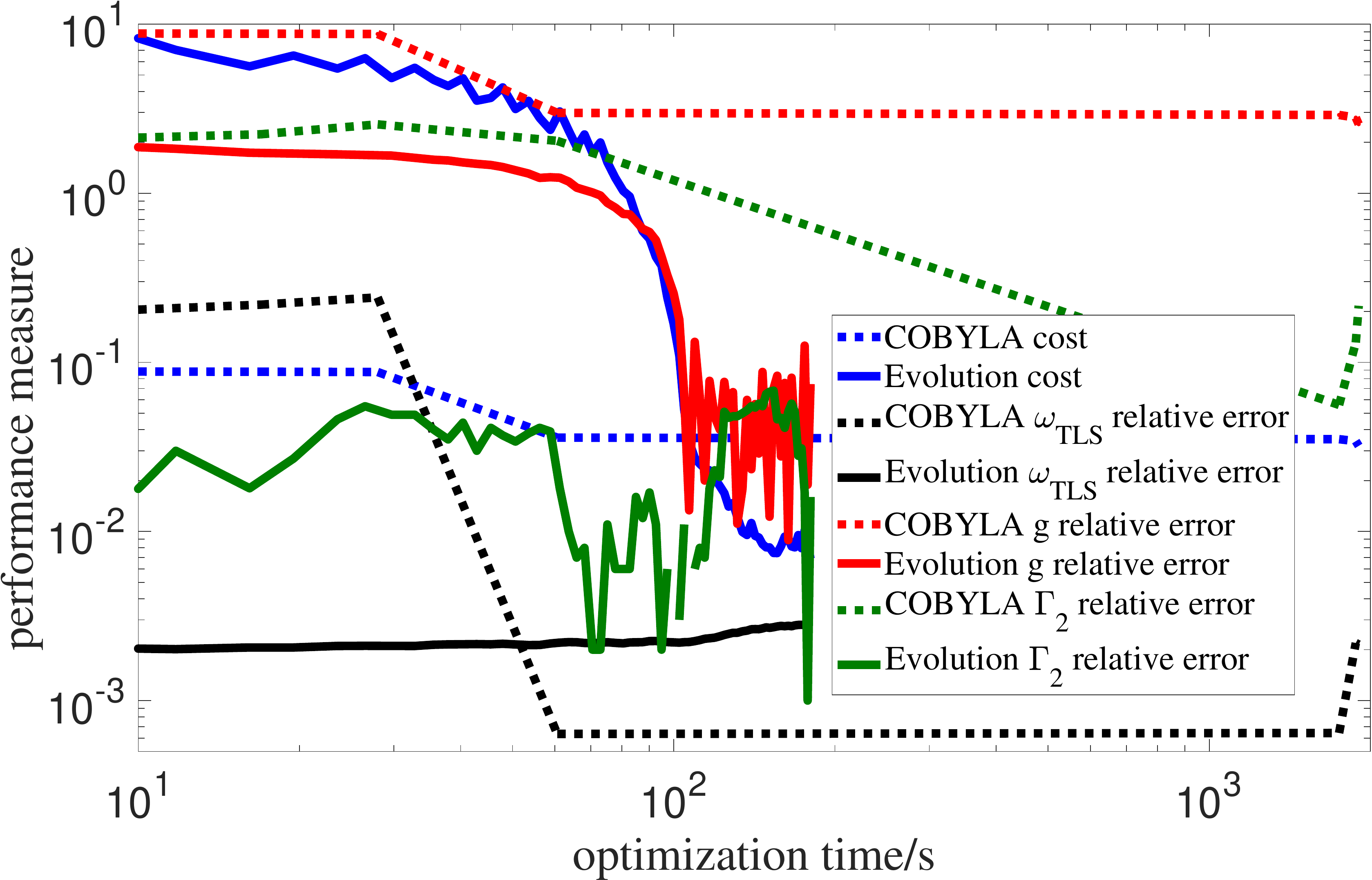}
\caption{Fitting performance comparison between DNN-EA and COBYLA optimizer. 
\label{CompareMethods}}
\end{center}
\end{figure}

We deploy DNN-EA for robust learning of the underlying  TLS model parameters: $\lambda =8.959$MHz, $\Gamma_2=10$MHz, $\ftls=5.04$GHz, $\tr =6$ns, and successfully reproduce the experimental data in all four different frequency regimes of swap spectroscopy see Fig.~\ref{sumplotfitexperiment}. 
Our DNN based method outperforms traditional optimizers and exhaustive search, in both the  fitting accuracy and the rate of convergence in the learning. Both exhaustive search and gradient based COBYLA methods can be blind-sighted by  false optimal points of the cost function due to experimental noise in the data. Our DNN-EA, in comparison, is   resilient against the spurious effect of noise and robustly find the globally optimal parameters.  Fig.~\ref{CompareMethods} summarizes   the performance comparisons of COBYLA optimizer and DNN-EA in regard to the cost defined in Eq.~(\ref{meansquareError})   and  the relative error for each TLS parameter defined by its difference from the correct value divided by the amplitude of each parameter. 
In particular, DNN-EA  achieves one magnitude lower L2 error in predicting the TSSD~(defined in Eq.~(\ref{meansquareError}) ) than traditional COBYLA optimizer with a wall-clock runtime of  around 100s. In contrast, a grid search on all the parameters takes $10^6$s for a common work station  which we use to perform the optimization.  And the relative error for each parameter of DNN-EA is consistently lower than that of COBYLA.

Next we address how average two-qubit-gate fidelity depends on the physical parameters of environmental defects and control waveforms. Our goal is to connect high-level digital circuit performance to low-level quantum physics of the device, which is essential for developing new characterization methods and error mitigation strategies  for  quantum computation. We derive an operator sum description of the non-Markovian quantum channel for the  two-qubit  gate, which is then used  to compute the average gate fidelity  as  a function of TLS model parameters and gate parameters~(details see  Supp.~\ref{SuppKraus}). Our model includes both Markovian errors due to electronic white noise  and non-Markovian errors due to TLS-qubit dynamics. 
 
The  most general form of a two-qubit gate realizable in Google's superconducting architecture~\cite{niu2019universal,Barends2019, QuantumSupremacy2019} can be defined as~(details see Supp.\ref{SuppKraus}):
\begin{align}
U_2
&=\left(\begin{matrix}1 & 0& 0& 0\\
0 & e^{-i \phi}\cos( \theta ) & -i \sin(\theta) e^{i \phi}& 0\\
0 & -i \sin(\theta) e^{i  \phi} &  e^{-i  \phi}\cos(\theta) & 0\\
0& 0& 0 & e^{-i 2 \phi + i\psi}  
\end{matrix}\right).
\end{align}    In hardware, such a gate may be realized via frequency control as follows. First, each qubit is detuned from it's respective idle frequency $f_{\text{idle}}$~(sufficiently far apart  from one another~\cite{martinis2014fast}) towards a common interaction   frequency $f_q$. The two-qubit interaction $g( \sigma_1^x\sigma_2^x + \sigma_1^y\sigma_2^y)$ is then actuated through a tunable coupler~\cite{Chen2014} for a gate time $t_{2}$ resulting in the wanted rotation $\theta = g t_2$. Finally, the interaction is turned off and each qubit is detuned back to it's respective $f_{\text{idle}}$. 
By setting  $t_{\text{p}}=t_2$ and \fp$=f_q$, each qubit executes a swap-spectrscopy gate. Consequently, the  degree of decoherence during   $U_2(g, t_2, f_q)$ is directly  related to  the qubit decay probability from TSSD.

\begin{figure*}[ht]
\begin{center}
\includegraphics[width=0.9\textwidth]{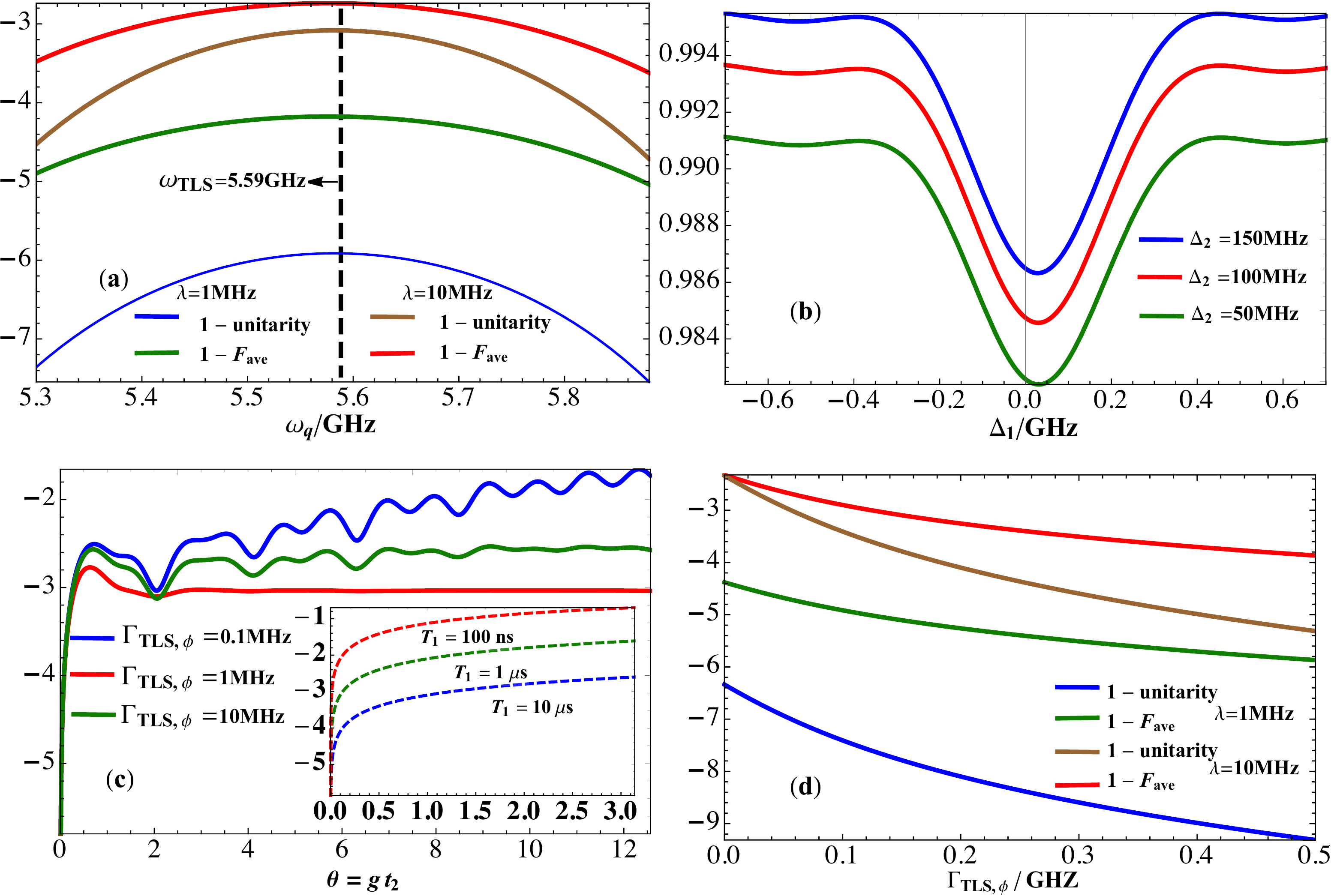}
\caption{ Two-qubit gate fidelity measures as a function of TLS parameters with:  $\lambda=10$MHz,   $g=50$MHz, $t_1=\pi/(4g)$,  $g_{\text{int}}=50$MHz.
For (a), (b) and (d), $t_{2}$ and $f_q$ are chosen to satisfy the condition for $U_2(g_{\text{int}}, t_2, f_q)=\sqrt{\text{ISWAP}}$~(see Supp.~\ref{FidelityEstimate}) with qubit idle frequency $f_{\text{idle}}=5.7$GHz.    (a): log scale average two-qubit gate error~(green and red curves) and $1- $ rescaled unitarity~(blue and brown curves) vs interaction frequency for different qubit-TLS interaction strength $\lambda$.
  (b): average two-qubit gate fidelity vs frequency of the first TLS measured by $\Delta_1=\omega_{TLS,1}-f_q$ for different value of $\Delta_2=\omega_{TLS,2}-f_q$ of the second TLS~(near the second qubit).
 (c): log scale average two-qubit gate error vs two-qubit gate rotation angle $\theta= g t_2$ under different TLS dephasing rate; inset: incoherent error due to qubit relaxation for a chosen range of coherence time.   (d): average two-qubit gate error~(green and red curves) and $1- $ rescaled unitarity~(blue and brown curves) vs TLS dephasing rate   for different qubit-TLS interaction strength $\lambda$. 
\label{2qubitgateFidelity_TLS_parameters}}
\end{center}
\end{figure*}

To derive the two-qubit gate error in a realistic setting~\cite{Klimov2018}, we focus on the case when one of the active qubits is operating near a TLS, while the other qubit is operating in a non-defective environment (see detailed derivation in Supp.~\ref{SuppKraus}).    Figs.~\ref{2qubitgateFidelity_TLS_parameters} (c) and (d) show that as the coupling strength between the TLS and qubit decreases, the incoherent contribution to the error budget, as measured by unitarity, decreases much faster than the coherent contribution. This signifies the importance of using a non-Markovian error model even in the presence of weakly coupled TLS.
   Our model also predicts that to maintain an average two-qubit gate error rate below $10^{-3}$, the qubits' common interaction frequency should stay at least $300$MHz away from TLS frequencies when the qubit-TLS coupling strength is $\sim 10$MHz (see Fig.\ref{2qubitgateFidelity_TLS_parameters}(a)).

With this newly established error model, we   discover three intriguing properties of  non-Markovian errors during a  two-qubit gate. First, a more coherent TLS  results in larger  gate error (see Fig.~\ref{2qubitgateFidelity_TLS_parameters} (c) and (d)). Second,  two-qubit gate error depends on the interaction frequency of the gate   symmetrically around the TLS frequency. Third, as the qubit-TLS interaction strength decreases, the incoherent error~(measured by rescaled unitarity) decreases much faster than the coherent error~(measured by the unitary gate error). 
This is observation is supported in Fig.~\ref{2qubitgateFidelity_TLS_parameters} (a) and (d), where the incoherent error is suppressed by more than three orders of magnitude while the overall average gate error is suppressed by only two orders of magnitude.  This clearly contrasts Markovian type errors, which are independent of the gate parameters and are worsened when the   TLS becomes more incoherent. These detailed error budget calculation   demonstrate that non-Markovian error dominates even in the weakly coupled limit when the qubits' interaction frequency is sufficiently  close to a TLS frequency. This debunks a widely used assumption that a Markovian TLS error model applies as long as the qubit-TLS coupling is sufficiently weak, and proves the importance of including non-Markovian models when characterizing TLS induced errors.

\section*{Conclusion}

 In this work, we advance the fundamental understanding of  the noise and decoherence of controllable quantum system by developing  physical models and efficient characterization scheme through machine learning to robustly infer the underlying model parameters from experimental data. An accurate characterization of quantum noise  is   critical   for the development of fault-tolerant salable quantum computers.  Towards this goal we first derive  a 
gate-dependent noise model in the perturbative and strongly coupled regimes   of qubit-TLS coupling during quantum gate operation.  Our   model's sensitivity to  qubits' frequency trajectories facilitates using qubits  as sensors to simultaneously characterize practical imperfections in both the qubit environment and  control electronics.
Combining this new physical model with novel machine learning algorithms and the Moir\'{e} effect, we  improve the accuracy and speed of learning TLS model parameters from noisy swap spectroscopy data by up to one order and two orders of magnitudes respectively. Lastly, we derive specific relations  between   physical model parameters and the  average gate errors induced by qubit-TLS coupling and frequency control imperfections. The learned physical parameters of TLSs can then directly be used to characterize and in turn  improve the performance of quantum gate.  Our results  represent an important step towards   \textit{in-situ}  quantum control optimization against both environmental and control defects.  The  realistic quantum noise channel description for two-qubit gate developed here also constitutes a missing piece of our understanding of realistic quantum devices. Our work is therefore   essential to the   simulation and analysis of the performance of noisy-intermediate scale processors in realizing gate-based quantum algorithms.   


\section*{Author contributions}
M. Niu, V. Smelyanskyi, and S. Boixo, developed the theoretical model, M. Niu developed and implemented the machine learning algorithm for the model inference. P. Klimov  carried out the experiment
and collected the data. The data analysis was done jointly
with the whole hardware team. All authors wrote and revised the
manuscript and supplement.

\onecolumngrid
\begin{appendix}

\section{Case I: Linear Perturbation Theory}
\label{SecLinearPerturbation}
In this section, we derive the decay probability of qubit to its ground state during a swap-spectroscopy gate. Here, we  use linear perturbation theory and assume that the coupling strength between TLS and qubit is weak enough for the Born approximation to apply. 
The Hamiltonian of the joint TLS-qubit system consists of three parts:
\begin{align}\label{EqTLSQUBITfirst}
&\hat{H}(t) =  \hat{H}_{TLS}(t) + \hat{H}_Q(t) + \hat{H}_{Q-TLS}(t)  
\end{align}
where $\hat{H}_{TLS}(t) $ includes the TLS Hamiltonian, its coupling to an environmental bosonic bath and the bath Hamiltonian:
\begin{align}
& \hat{H}_{TLS} = \hat{H}_{TLS,0} +  \hat{H}_{TLS-B_{TLS}} + \hat{H}_{B_{TLS}},\\
&\hat{H}_{TLS, 0}=- \frac{\hbar \omega_{TLS}}{2}\sigma_{TLS}^z - \frac{\hbar \epsilon_{TLS}}{2}\sigma_{TLS}^z,\\
& \hat{H}_{TLS-B_{TLS}} = \lambda_{TLS} (\sigma_{TLS}^+\hat{a}_B + \sigma_{TLS}^-\hat{a}_B^\dagger),\\
& \hat{H}_{B_{TLS}} = (\frac{1}{2} + \hat{a}^\dagger_{B}\hat{a}_{B})\hbar \omega_{B_{TLS}}
\end{align}
where   $B_{TLS}$   represents the TLS's bath, and $B_Q$ represents the  qubit's bath. Similarly, $ \hat{H}_Q(t)$ consists of qubit Hamiltonian,    qubit-bath coupling Hamiltonian, and qubit's bath Hamiltonian:
\begin{align}
& \hat{H}_Q(t) = \hat{H}_{Q,0}(t) + \hat{H}_{Q-B }(t) + \hat{H}_{B_Q}(t),\\
& \hat{H}_{Q,0}(t)  = - \frac{\hbar\epsilon(t)}{2}\sigma_Q^z,\\
&\hat{H}_{Q-B}(t) = \lambda_Q (\sigma_{Q}^+\hat{b}_B + \sigma_{Q}^-\hat{b}_B^\dagger),\\
& \hat{H}_{B_{Q}} = (\frac{1}{2} + \hat{b}^\dagger_{B}\hat{b}_{B})\hbar \omega_{B_{Q}}
\end{align}
where we consider the transversal  qubit-TLS interaction
\begin{align}\label{EqTLSQUBITlast}
& \hat{H}_{Q-TLS} = \lambda  \sigma^x_{TLS}\sigma^x_Q
\end{align}
with a interaction strength $\lambda$ which is weak enough compared to the range of interaction time $t$ of interest: $\lambda t\ll 1$. Notice that the longitudinal coupling does not directly affect the  the qubit decay rate  in the computational basis. However, we indirectly accounted for the effect of longitudinal coupling in altering the dephasing rate of the qubit which in turn alters the decay probability.

We move into the interaction picture, where the qubit and TLS operators can be expressed as
\begin{align}\label{interactionPicQubitOp}
& \tilde{ \sigma}_{Q}^{a} = \text{Tr}_{B_Q}\left[ \mathcal{T}_-\left[ \exp[i \int_0^t \hat{H}_{Q}(\tau)d\tau\right] \sigma_Q^a(0) \mathcal{T}_+\left[ \exp[-i \int_0^t \hat{H}_{Q}(\tau)d\tau\right] \right]\\\label{interactionPicTLSOp}
&\tilde{ \sigma}_{TLS}^a = \text{Tr}_{B_{TLS}}\left[ \mathcal{T}_-\left[ \exp[i \int_0^t \hat{H}_{TLS}(\tau)d\tau\right] \sigma_{TLS}^a(0) \mathcal{T}_+\left[ \exp[-i \int_0^t \hat{H}_{TLS}(\tau)d\tau\right] \right]
\end{align}
for $a \in \{x,y,z\}$. In this basis, the density operator for the joint system evolves with time as
\begin{align}
\tilde{\rho}(t) = -\frac{i}{\hbar}\int_{0}^t [\hat{H}_{Q-TLS}(s),\tilde{\rho}(s)] ds +\tilde{\rho}(0), \,\, \text{with}\,\, \tilde{\rho}(0) = \rho(0)
\end{align}
To the second order perturbation theory, the above differential equation can be solved as 
\begin{align}
\tilde{\rho}(t) &= \rho(0) - \frac{i}{\hbar}\int_0^t[ \hat{H}_{Q-TLS}(s), \rho(0)] ds - \frac{1}{\hbar^2}\int_0^t ds \int_0^s d \tau [\hat{H}_{Q-TLS}(S), [\hat{H}_{Q-TLS}(\tau), \rho(0)]] + O[(gt)^3]\\\label{blochRedfield11}
&= \rho(0) - \frac{ig}{\hbar}\int_0^t \left[ \tilde{\sigma}_Q^x(s)\tilde{\sigma}_{TLS}^x(s) \rho(0) - \rho(0) \tilde{\sigma}_Q(s)\tilde{\sigma}_{TLS}(s) \right] ds 
\end{align}
 Now we apply the Born approximation assuming that the qubit and   TLS remain  in a product state given their  initial state as a product state $\rho(0) =\rho_Q(0)\otimes\rho_{TLS}(0)= \ket{1}\bra{1}_Q \otimes \ket{0}\bra{0}_{TLS}$.  Inserting this condition into Eq.~(\ref{blochRedfield11}) above while tracing out the TLS system gives us the evolution of the qubit density operator as:
\begin{align} \label{qubitdensityOpEQ}
\tilde{\rho}_Q(t) &=\rho_Q(0) - \frac{\lambda^2}{\hbar^2}\int_0^t ds \int_0^s d \tau\left[ C(s-\tau)\tilde{\sigma}_{Q}^x(s)\tilde{\sigma}_{Q}^x(\tau)\rho_Q(0)  + C(\tau-s)\tilde{\sigma}_{Q}^x(\tau)\tilde{\sigma}_{Q}^x(s)\rho_Q(0)    \right.\\\nonumber
&\left.- C(\tau-s) \tilde{\sigma}_{Q}^x(s)\rho_Q(0) \tilde{\sigma}_{Q}^x(\tau)  - C(s-\tau) \tilde{\sigma}_{Q}^x(\tau)\rho_Q(0) \tilde{\sigma}_{Q}^x(s)] \right]
\end{align} 
where we use $C(\tau -s) = \langle \tilde{\sigma}_{TLS}^x(\tau)  \tilde{\sigma}_{TLS}^x(s)\rangle$ to represent the correlator of TLS.  Solving  the dynamical evolution of TLS under its coupling to a Markovian environment separately~\cite{Shnirman2003} gives us $C(\tau -s) =   e^{-\Gamma_{TLS, \phi} |\tau - s| - i \omega_{TLS}(\tau-s)} $,
with $\Gamma_{TLS, \phi}=\frac{\Gamma_l^{TLS}}{2}+\frac{2\alpha k_B T}{\hbar}$ representing the dephasing rate of TLS caused by longitudinal coupling to an Ohmic environment at temperature $T$ and transversal coupling~($\sigma^z_{TLS}$) to the environment. We note, however, that the Markovian assumptions for the TLS's environment  are not fundamental. By replacing  the current TLS  correlator with that from a non-Markovian environmental couplings,   the following derivations still apply.   

We now investigate the qubit operator in the interaction picture under the coupling to a separate Markovian environment.  After tracing out the qubit's   environment  in Eq.~(\ref{interactionPicQubitOp}), we obtain the two-qubit operator:
\begin{align}
\langle\tilde{\sigma}^x_Q(\tau)\tilde{\sigma}^x_Q(s)\rangle =\left(\cos\phi(\tau)\sigma_Q^x +\sin\phi(\tau)\sigma_Q^y \right) \left(\cos\phi(s)\sigma_Q^x +\sin\phi(s)\sigma_Q^y \right)  e^{-\Gamma_{2,q} \vert \tau -s \vert }   
\end{align}
with the dynamical phase   accumulated during  the swap-spectroscopy gate defined by:
\begin{align}\label{pulseshapeQubitEq}
\phi(\tau) = f_{\text{idle}}\tau + \int_0^\tau \epsilon_Q(t) dt =  f_{\text{idle}}\tau - \epsilon_{\text{max}} \int_0^\tau \mu(t) dt.
\end{align}
We use $\mu(t)\in[0,1]$ as a unitless time-dependent function rescaled by the maximum frequency change   $\epsilon_{\text{max}}$ during the swap spectroscopy and the minus sign in front of $\mu(t)$ is due to our experimental convention where most of the TSSD is taken by lowering the qubit frequency from its initial value.
Inserting these results into Eq.~(\ref{qubitdensityOpEQ}) while adopting   the Rotating Wave Approximation~(RWA) gives us the qubit density operator in the interaction picture at the end of swap-spectroscopy gate:
\begin{align}
\tilde{\rho}(t_{\text{p}}) = \ket{1}\bra{1}_q +   \frac{\lambda^2}{\hbar^2}\sigma_Q^z(0) \int_0^{t_{\text{p}}} ds \int_0^s d\tau e^{- \Gamma_{2} |\tau- s| } \left(  e^{-i (\phi(\tau)- \phi(s) - \omega_{TLS}(\tau-s))} + e^{i (\phi(\tau)- \phi(s) - \omega_{TLS}(\tau - s)}  \right)
\end{align}
where $t_{\text{p}}$ represents the total duration of the swap spectroscopy gate,  the exponential decay factor $ \Gamma_{2}= \Gamma_{TLS,\phi} + \Gamma_{2,	q}$   is contributed both from qubit's own coupling to the environment~($1/\Gamma_{2,	q}$ is of order 5 $\mu$s) and indirectly through the coupling of TLS its own environment. 
Projecting the qubit density operator above onto the ground state gives us  qubit's decay probability    after the swap spectroscopy gate:
\begin{align}\nonumber
P_e & = \bra{0}_q\tilde{\rho}_q(t_{\text{p}})\ket{0}_q \\\nonumber
&=\tilde{\lambda}^2\text{Re}\left[  \int_0^T dx\int_0^Tdy e^{-\tr\Gamma_2 |x-y|} e^{ i   \epsilon (x-y)+ \eta(\Phi(x)- \Phi(y))} \right] ,\\\label{decayProbBeforeSPA}
&= \tilde{\lambda}^2 R(T, \gamma, \epsilon, \eta) \\&R(T, \gamma, \epsilon, \eta)=   \text{Re}\left[ \int_0^T dx \int_0^T dy e^{-\gamma\vert x-y \vert}e^{iF(x,y)}\right]
\end{align} and  we chose   unit-less parameters $\tilde{\lambda}=\lambda \tr/\hbar, \gamma = t_r \tilde{ \Gamma}_2, \epsilon = \tr (f_{\text{idle}} -\omega_{TLS}), \eta = \tr \epsilon_{\text{m}}, \Phi(x) = \int_0^x \mu(x)dx$, and unit-less time $x $ and $ y$  normalized   by the ramp time  $t_r$  of the qubit frequency control trajectory~(see Fig.~\ref{PulseShapeSwapspec}). This  gives the rescaled  overall runtime $T= t_{\text{p}}/\tr$. We   use   $F(x,y)=  \epsilon (x-y) -\eta(\Phi(x)- \Phi(y))$ to represent the dynamical phase accumulated during the gate.  This dynamical phase   under the frequency trajectory of Fig.~\ref{PulseShapeSwapspec}   obeys the following   relations:
\begin{align}\label{pulseunitlessEq}
\frac{\epsilon_Q(t) }{\epsilon_{\text{max}}}=- \mu\left(\frac{t}{t_r}\right), \quad \mu(0)=\mu(T)=0, \quad \mu(1)= \mu(T-1),\\\label{phaseunitlessEq}
\Phi(x)= \int_0^x \mu(\tau) d\tau,\quad \Phi(0)=0,\quad \Phi(T-1)=\Phi(T)- \Phi(1),
\end{align}
where the last two equations of  both lines are given by the time reversal symmetry of the   qubit control pulse shape.

 Numerical  integration of  Eq.~(\ref{decayProbBeforeSPA})   proves to be too slow to be  suitable for fast online TLS characterization. We notice that during the ramp time, the phase   $\Phi(x)$ oscillates at a unitless speed of  $\frac{d \Phi(x)}{dx} =  t_r\epsilon_{max}\approx 6\text{ns}\times 2\pi \times 0.5\text{GHz}\approx 20\gg 1 $, which is fast oscillating. This means, the integration contributed from   the ramp up $\int_0^1$ and ramp down $\int_{T-1}^T$ can be approximately calculated using a popular method in path integral: stationary phase approximation.
 
When integrating a fast oscillating function in part of Eq.~(\ref{decayProbBeforeSPA}), according to the stationary phase approximation the main contribution to the integral are from   the stationary points $x_c, y_c$ when the gradient of the phase vanishes:
\begin{align}\label{StationaryConditionEq1}
&\partial_x  F(x, y)|_{x=x_c, y = y_c}=\epsilon  - \eta \mu(x)|_{x=x_c, y = y_c}  =t_r[f_{\text{idle}}-\epsilon_{\text{max}}\mu(x_c)+ \omega_{tls} ]=0,\\\label{StationaryConditionEq2}
& \partial_y F(x, y)|_{x=x_c, y = y_c}=-\epsilon+ \eta \mu(y)|_{x=x_c, y = y_c} =-t_r[f_{\text{idle}}-\epsilon_{\text{max}}\mu(y_c)+ \omega_{tls} ]=0.
\end{align}
The stationary condition is met whenever the qubit frequency reaches the TLS frequency during the ramping up $x\in[0,1]$ or ramping down $x\in[T-1, T]$ of frequency control.  We  separate the double integration in Eq.~(\ref{decayProbBeforeSPA}) into nine parts according to:
\begin{align}
&\mathcal{T}_1=[0,1),\quad \mathcal{T}_2 =[1, T-1),\quad \mathcal{T}_3 =[T-1, T].\\
&R(T, \gamma, \epsilon, \eta)= \sum_{\alpha, \beta=1}^3R_{\alpha, \beta},\quad R_{\alpha, \beta}= \int_{x\in \mathcal{T}_\alpha} dx \int_{y\in \mathcal{T}_\beta} dy e^{-\gamma |x-y|} e^{i F(x,y)}
\end{align}
 where we use $\mathcal{T}_1$ and $\mathcal{T}_2$ to represent the first and second ramps, and $\mathcal{T}_3$ to represent   plateau  where qubit frequency is fixed at   $\fp = f_{\text{idle}} - \epsilon_{\text{max}}$. Each one of the nine parts is derived separately as follows.
 
 \begin{enumerate}
 \item $R_{22}$: when both integrals are in the region of a constant   frequency at the plateau, we have closed-form expression: 
  
 \begin{align}\nonumber
 \text{Re}\left[ \int_{1}^{T-1} dx \int_{1}^{T-1} dy e^{-\gamma |x-y|} e^{i F(x,y)}\right]=&\frac{2  (-\gamma ^2+\gamma  (T-2)  (\gamma ^2+(\epsilon -\eta
  )^2 )+e^{\gamma  (2-T)} (\left(\gamma ^2-(\epsilon -\eta )^2\right) \cos
  ((T-2) (\epsilon -\eta ))}{\left(\gamma ^2+(\epsilon -\eta )^2\right)^2} \\
  &  +\frac{2 \gamma  (\eta -\epsilon ) \sin ((T-2) (\epsilon -\eta
  )) +(\epsilon -\eta )^2 )  }{\left(\gamma ^2+(\epsilon -\eta )^2\right)^2}
 \end{align}
 
 \item $R_{11}$: when both integrals are over the ramp-up part of the swap-spectroscopy gate,  during which the stationarity condition in Eq.~(\ref{StationaryConditionEq1}) and (\ref{StationaryConditionEq2}) are met at $x_c$. We  expand the  unitless phase in Eq.~(\ref{phaseunitlessEq}) around this stationary point:
 \begin{align}
  \Phi(x) \approx  \frac{1}{2} x_c
  \left(x-x_c\right)^2 f^{'}+\frac{\epsilon  \left(x-x_c\right)}{\eta }+\Phi(x_c)\\
    \Phi(y) \approx  \frac{1}{2} x_c
  \left(y-x_c\right)^2 f^{'}+\frac{\epsilon  \left(y-x_c\right)}{\eta }+\Phi(x_c)
 \end{align}
 which gives a simplified overall phase factor:
 \begin{align}\label{EqFapproximate}
 F(x,y) \approx -\frac{\eta f^{'}(x_c)}{2}\left((x-x_c)^2 - (y-x_c)^2\right)
 \end{align}
 which gives 
 \begin{align}
 R_{11} &=\frac{2\pi}{\eta \vert f^{'}(  x_c) \vert}\left\vert \text{erf}(z_1, z_2)\right\vert^2
 \end{align} where we use simplified notation: $z_1= - e^{i\pi/4}(1-x_c)\sqrt{\frac{\eta \vert f^{'}(  x_c) \vert}{2}},  $ and $z_2=  e^{i\pi/4} x_c\sqrt{\frac{\eta \vert f^{'}(  x_c) \vert}{2}}$.
 
  \item $R_{33}$: due to the symmetry described in Eq.~(\ref{phaseunitlessEq}) and (\ref{pulseunitlessEq}), we have $R_{33}=R_{11}$.
  
  \item $R_{13}$ and $R_{31}$:  using similar expansion around the stationary point as for $ R_{11} $ and the symmetry argument in Eq.~(\ref{phaseunitlessEq}) and (\ref{pulseunitlessEq}), and under the same notation of $z_1, z_2$: 
  \begin{align}
 R_{13}=R_{31}= \frac{\pi  e^{-\gamma  \left| T-2 x_c \right| }
  \text{Im}\left[\text{erf}\left(-z_2, -z_1\right)^2 \exp \left(i \eta  \left(T \phi -2 \phi   x_c\right)-i \epsilon  \left(T-2 x_c\right)\right)\right]}{2 \eta   f^{\prime}(x_c)}
  \end{align}
  
  \item $R_{12}$ and $R_{21}$: one of the integral is in ramp-down part and the other is in plateau part of the frequency trajectory. We expand one of the double integrals around the stationary point to obtain the analytic integration result:
  \begin{align}\nonumber
  R_{12}& =\text{Re}\left[ \int_{0}^{1} dx \int_{1}^{T-1} dy e^{-\gamma |x-y|} e^{i F(x,y)}\right],\\\nonumber
  &\approx\frac{\sqrt{2 \pi } e^{\gamma  x_c}
 \left\vert\left[\text{erf}\left(z_2\right)-\text{erf}\left(z_1\right)\right]\right\vert \sqrt{\frac{1}{\eta  f^\prime( x_c)}}}{\sqrt{\gamma^2+(\epsilon -\eta  )^2}}
 \left| e^{-\gamma }-e^{\gamma  (1-T)-i T (\epsilon -\eta )}\right|  \\
 &\times
 \cos \left(-\arg(\gamma -i (\epsilon -\eta ))-\arg \left(e^{-\gamma }-e^{\gamma  (1-T)-i
  T (\epsilon -\eta)}\right)\right.\\\nonumber
  & \left. -\arg\left[\left[\text{erf}\left(z_2\right)-\text{erf}\left(z_1\right)\right]\right]-\epsilon  \left(x_c-1 \right)+\eta \phi  x_c-\eta (\phi +2)-\pi /4 \right),
  \end{align}here and henceforth, the approximation similar to Eq.~(\ref{EqFapproximate}) are used. 
Due to the symmetry of the qubit dynamics around the half evolution time, we also have $R_{21}=R_{12}$.
 
 \item $R_{23}$ and $R_{32}$: one of the double integrals is in plateau part of the frequency trajectory, and the other one is in the ramp-up part of the pulse. After expanding the ramp-up part around the stationary point we have:
 \begin{align}
 R_{23}&= \text{Re}\left[ \int_{1}^{T-1} dx \int_{T-1}^{T} dy e^{-\gamma |x-y|} e^{i F(x,y)}\right],\\\nonumber
 &\approx \frac{\sqrt{\frac{\pi }{2}} \left(e^{(T-1) (\gamma +i (\epsilon -\eta
  ))}-e^{\gamma +i (\epsilon -\eta )}\right) \left(\text{erf}\left(\frac{x_c \sqrt{i
  \eta  f^\prime\left(x_c\right)}}{\sqrt{2}}\right)-\text{erf}\left(\frac{\left(x_c-1\right)
  \sqrt{i \eta  f^\prime\left(x_c\right)}}{\sqrt{2}}\right)\right) }{(\gamma +i (\epsilon -\eta )) \sqrt{i \eta 
  f^\prime\left(x_c\right)}} \\
  &\times \exp \left(-\gamma 
  \left(T-x_c\right)+i \left(\varepsilon  \left(x_c-T\right)+\eta \left(-\phi  x_c+T
  \phi -\phi +1\right)\right)\right)
 \end{align}
 \end{enumerate}

The   decay probability after the swap-spectroscopy gate is the weighted  sum of all nine terms derived above: $P_{\text{decay}}=\left(\frac{g \tr}{\hbar}\right)^2\sum_{\alpha,\beta=1}^3R_{\alpha,\beta}$. This close form expression reduces the evaluation time of TSSD cost funciton, which consists of the sum of many $P_{\text{decay}}$ data points, by three magnitude from direct numerical integration.  The applicability of stationary point approximation depends on the magnitude of $\eta$ which determines how fast the phase in the integration oscillates. In our case as estimated before $\eta\approx  20$, which gives an approximation  error compared to the exact numerical simulation is numerically bounded to below $10^{-3}$.

\section{Case II: Coherent TLS-qubit Interaction}\label{SecCoherentTLSqubit}
In this section, we derive the   coherent TLS-qubit coupling model under two different qubit frequency control trajectories. Based on these results, we analyse the conditions when the gate-dependency of the qubit-TLS interaction dynamics is non-negligible.

The transversal coupling between TLS and qubit conserves the total excitations. Since qubit is always initialized in the excited state, the subspace of quantum dynamics of TLS and qubit is spanned by two orthogonal states: $\{\ket{1,0}, \ket{0,1}\}$, where we use  the first number in the ket representation as the excitation number in TLS mode. The case when the initial state of TLS is in  excited state is neglibible due to  two important facts: 1. TLS is much less coherent than qubit and decay quickly to its ground state, and is thus less likely to be in excitated state; 2. population transfer is unaffected if TLS and qubit both starts in excited state due to the excitation conserving nature of the qubit-TLS coupling.  The qubit-TLS Hamiltonian in this single-excitation subspace  takes the following form:
\begin{align}
\hat{H}_{TLS,q}(t)=\left(\begin{matrix}
\frac{\Delta(t)}{2} & \lambda\\
\lambda & -\frac{\Delta(t)}{2}
\end{matrix}\right),
\end{align}
where $\Delta(t)=f_q- \omega_{TLS}$ represents the energy gap between the TLS and qubit energy levels, and $g$ is the TLS-qubit coupling strength. 

During  swap spectroscopy, the energy gap changes according to a smoothed-trapezoidal function for time $T$ and induces the unitary transformation on the joint system according to $U(T)=\mathcal{T}[\exp[-i \int_0^T\hat{H}_{TLS,q}(t) dt]$. 

We   decompose the overall unitary into a product of three parts. The first part is induced by a time-dependent Hamiltonian, where the energy gap $\Delta(t)$ changes  from an initial value $\Delta_0$ to a plateau value $\Delta_i$ during the rise time $t_r $ as $U_1= \mathcal{T}[\exp[-i \int_0^{t_r} \hat{H}_{TLS,q}(t) dt]$. The second part is the unitary evolution under  a constant Hamiltonian  with $\Delta(t)=\Delta_i$ for time $t_{\text{p}}$ as: $U_2 = e^{-i\omega_1 t_{\text{p}}}\ket{\lambda_1}\bra{\lambda_1} + e^{-i\omega_2 t_{\text{p}}}\ket{\lambda_2}\bra{\lambda_2}$ with $\ket{\lambda_i}$ being the $i$th eigenvector of energy $\omega_i$. This is represented by the plateau part of the trapezoidal pulse in Fig.~\ref{PulseShapeSwapspec}.

The third part $U_3$ corresponds to the unitary evolution under a time-dependent Hamiltonian,  which is a time reversed version of $U_1$. The overall decay probability depends on the three unitaries as
\begin{align}
P_{decay}&= \vert \bra{1,0}U_3 U_2 U_1 \ket{0,1}\vert^2
\end{align}

\subsection{Rectangular Pulse}\label{recPulseSec}

Having a rectangular frequency control trajectory in place of a trapezoidal pulse amounts to taking the limit of $U_1\to I$ and  $U_3\to I$. In this case,  the probability of transition is   purely contributed from the plateau part by a a rectangular pulse $P_{rec}$:
\begin{align}
P_{rec} =\vert \bra{1,0}  U_2   \ket{0,1}\vert^2= \frac{4\lambda^2}{4\lambda^2 + \Delta_1^2}\sin\left(\frac{1}{2}t_{\text{p}} \sqrt{4\lambda^2 + \Delta_1^2}\right)^2
\end{align}

\subsection{Trapezoidal Pulse}\label{TrapPulseSec}

If we represent the unitary during ramp up and ramp down by:
\begin{align}
&U_1=\left(\begin{matrix}
u_{11}  & - u_{21}^*\\
u_{21}  & u_{11}^*
\end{matrix}\right)\\
  &  U_3=\left(\begin{matrix}
w_{11} & - w_{21}^*\\
w_{21}  & w_{11}.
\end{matrix}\right)
\end{align}
We find each element of these two unitary matrices can be found by solving the Schr\'{o}dinger equation under a linearly changing Hamiltonian:
\begin{align}\label{u21Eq}
&u_{21}= -\frac {i s (\nu + \xi [s])^{3/
      2}\Gamma [-\nu - \xi [s]]} {(\nu \xi [s] + 
      1)\sqrt {2\pi}}\sum_{\alpha\in {-1, 1, 2}}\alpha PC[\nu -  s, -\alpha z_{in}] PC[\nu - s, \alpha z] \\\label{u21Eq2}
      & w_{21}= -\frac {i s (\nu + \xi [s])^{3/
      2}\Gamma [-\nu - \xi [s]]} {(\nu \xi [s] + 
      1)\sqrt {2\pi}}\sum_{\alpha\in {-1, 1, 2}}\alpha PC[\nu -  s, \alpha z] PC[\nu - s, -\alpha z_{in}]\\
      &u_{11}=-\frac{(\nu +\xi (s)) \Gamma (-\nu -\xi (s)) (D_{\nu }(z) D_{\nu -s}(-\text{z}_{in})+D_{\nu }(-z) D_{\nu -s}(\text{z}_{in}))}{\sqrt{2 \pi }}\\
      &w_{11}=-\frac{(\nu +\xi (s)) \Gamma (-\nu -\xi (s)) (D_{\nu }(-\text{z}_{in}) D_{\nu -s}(z)+D_{\nu }(\text{z}_{in}) D_{\nu -s}(-z))}{\sqrt{2 \pi }}
\end{align}
with $PC$ stands for parabolic cylinder function and other variables defined as:
\begin{align}
  &  s=\text{sign}[v], \quad \xi =(1-s)/2, \quad \nu= -i \frac{\lambda^2}{4v}-\xi(s)\\
    &z_{in}=e^{i\pi/4}\frac{ \Delta_0 }{\sqrt{v}},\quad z =e^{i\pi/4}\frac{ \Delta_1  }{\sqrt{v}}.
\end{align}
 
\subsection{Gate-dependence of  TSSD}

The shape of the quantum gate control trajectory can be important to the qubit-TLS dynamics   under a non-Markovian TLS-qubit coupling.  The main difference between a rectangle pulse   analyzed in Sec.~\ref{recPulseSec} and that of a trapezoidal pulse analyzed in Sec.~\ref{TrapPulseSec}  lies in two parts: first only the trapezoidal pulse admits  Landau-Zener~(LZ) transition due to the finite velocity when crossing the   TLS frequency; second, the phase accumulated during the ramp up and ramp down will change the overall oscillation phase of the  decay probability as a function of ramp time.
To dissect the gate-dependence of TSSD,  we  analytically compare these two different  contributions from the time-dependent   frequency control trajectory.
We start by first simplifying our analytic  solution under a trapezoidal pulse by taking  the limit of $\delta =\lambda^2/(4v) \ll 1$. We analyze the ramp up unitary first,  then the ramp-down unitary, and finally compare their respective contributions to the   Landau-Zener-Rabi oscillation.

\subsubsection{Ramp up}\label{RampupSec}
In the first part of trapezoidal pulse, for time $0\leq t \leq t_{\text{r}}$ in Fig.~\ref{PulseShapeSwapspec} of the main text,  we have $s=1, \xi[s] =0, \nu = -i\delta$,  $\nu \xi[s]+1=1$, and $\nu + \xi[s] =-i\delta$. In this case the Gamma function can be expanded around $\delta$ as:
\begin{align}
\Gamma [-\nu - \xi [s]]=\Gamma [i\delta]=\frac{1}{12} i \left(6 \gamma ^2+\pi ^2\right) \delta +\frac{1}{i\delta }-\gamma + O(\delta^3)
\end{align}
and the coefficient for the parabolic cylinder function can be expanded around the small value $\nu = - i \delta$ as:
\begin{align}
\frac {i s (\nu + \xi [s])^{3/ 2}\Gamma [-\nu - \xi [s]]} {(\nu \xi [s] + 1)\sqrt {2\pi}} \approx \frac {i (-i\delta)^{3/ 2}} {\sqrt {2\pi}} (-\frac{1}{-i\delta }-\gamma + O(\delta^2))=\frac {i } {\sqrt {2\pi}} (-\sqrt{-i\delta}  + O(\delta^{3/2})) 
\end{align}
where the Euler constant is represented by $\gamma=0.57721...$. The leading contribution of the off-diagonal term of the ramp up unitary is thus of the magnitude $O(\sqrt{\delta})$. The other parts of the off-diagonal term in Eq.~(\ref{u21Eq}) can be also simplified using the approximation under $\nu -s \approx -1$:
\begin{align}
PC[\nu -s, x]\approx PC[-1, x]= e^{x^2/4}\sqrt{\frac{\pi}{2}}\text{erfc}\left(\frac{x}{\sqrt{2}}\right).
\end{align}
Under this approximation, we can rewrite the the off-diagonal part of  the ramp up unitary as:
\begin{align}\label{EqW21}
w_{21} &=-\frac{\sqrt[4]{-1} g e^{\frac{i \left(\Delta _0^2+\Delta _1^2\right)}{4 v}}
  \left(\text{erf}\left(\frac{\left(\frac{1}{2}+\frac{i}{2}\right) \Delta
  _0}{\sqrt{v}}\right)+\text{erf}\left(\frac{\left(\frac{1}{2}+\frac{i}{2}\right) \Delta
  _1}{\sqrt{v}}\right)\right)}{\sqrt{v}}
\end{align}
 where we can further simplify it by  the error function: $|\text{erf}(x)| \approx |x|$ when $|x|\ll 1$ as
 \begin{align}
    |w_{21} |  & \approx \left\vert \frac{\lambda (\Delta_0 +\Delta_1)}{v}\right\vert.
 \end{align}

\subsubsection{Ramp down}\label{RampDownSec}
In the third part of a trapezoidal trajectory, for time $\tr + \tp \leq  t \leq 2\tr + \tp  $ in Fig.~\ref{PulseShapeSwapspec}, we have $s=-1, \xi =1, \nu =-i \delta -1 $,  with the Gamma function $\Gamma[i\delta]$ sharing the same expansion as above, but $\nu \xi[s]+1=-i\delta$, and $\nu + \xi[s] =-i\delta$,
and the coefficient for the parabolic cylinder function can be simplified to the leading order  as
 \begin{align}
&\frac {i s (\nu + \xi [s])^{3/ 2}\Gamma [-\nu - \xi [s]]} {(\nu \xi [s] + 
      1)\sqrt {2\pi}} \approx \frac{-i(-i\delta)^{3/2}}{\sqrt{2\pi}(-i\delta)} \left(+\frac{1}{i\delta }-\gamma\right)= \frac{-i}{\sqrt{2\pi} } \left(-\frac{1}{(-i\delta)^{1/2} }-\gamma(-i\delta)^{1/2}\right)\\
      & = \frac{i(1 - i \gamma \delta) }{\sqrt{2\pi} \sqrt{-i\delta}} + O(\delta^{3/2})=\frac{i }{\sqrt{2\pi} \sqrt{-i\delta}}+ O(\delta^{1/2})
\end{align} 
The   off-diagonal term in Eq.~(\ref{u21Eq}) can be   simplified using the approximation under $\nu -s \approx  1$ similar to that  in the derivation of $w_{21}$  as
\begin{align}
u_{21}&= -\frac{(-1)^{3/4} g e^{-\frac{i \left(\Delta _0^2+\Delta _1^2\right)}{4 v}} \left(-i
  \text{erfi}\left(\frac{\left(\frac{1}{2}+\frac{i}{2}\right) \Delta _0}{\sqrt{\left| v\right| }}\right)-i
  \text{erfi}\left(\frac{\left(\frac{1}{2}+\frac{i}{2}\right) \Delta _1}{\sqrt{\left| v\right| }}\right)\right)}{\sqrt{\left|
  v\right| }}\\
  |u_{21}|& \approx \left\vert \frac{g (\Delta_0 +\Delta_1)}{v}\right\vert
\end{align} 
where the second linear approximation applies when $g (\Delta_0 +\Delta_1)/v\ll 1$. This simplification shows that the amplitudes of the ramp up and ramp down off-diagonal elements are identical $|u_{21}| =|w_{21}|$ in the large velocity limit.

\subsection{Landau-Zener-Rabi Oscillation}\label{LZdiscussionSec}
The whole swap spectroscopy gate  is a product of ramp up~$U_1$, plateau~$U_2$ and ramp down~$U_3$ unitaries.  We   express the transition  probability between $\ket{1,0}$ and $\ket{0,1}$ due to Landau-Zener-Rabi oscillation with   the matrix elements of $U_1$,  $U_2$ and $U_3$ as:
\begin{align}
P_{LZR}&= \vert \bra{1,0}U_3 U_2 U_1 \ket{0,1}\vert^2,\\
&=\left\vert \left(\begin{matrix}
1 & 0
\end{matrix}\right)\left(\begin{matrix}
w_{11} & - w_{21}^*\\
w_{21}  & w_{11}
\end{matrix}\right) \left(\begin{matrix}
v_{11} & - v_{21}^*\\
v_{21} & v_{11}^*
\end{matrix}\right)\left(\begin{matrix}
u_{11}  & - u_{21}^*\\
u_{21}  & u_{11}^*
\end{matrix}\right) \left(\begin{matrix}
0\\
1
\end{matrix}\right)\right\vert^2\\
&= \vert -w_{11}  \left(v_{11}u_{21}^* + u_{11}^* v_{21}^*\right) + w_{21}^*\left( v_{21} u_{21}^* - u_{11}^* v_{11}^* \right) \vert^2 
\end{align}

 The contributions to the overall transition probability from ramp up and down parts of the unitary evolution amounts to  adding a phase to the plateau part rabi oscillation $U_2$ by     $\theta$. To   the leading order, it depends on the coupling strength, energy gap and the velocity as:
\begin{align}
\theta = 2\frac{g|\Delta|}{v},
\end{align} where $\Delta=\Delta_0+\Delta_1$ and a factor of 2 comes from the individual contribution from ramp up and ramp down.   Since the rectangular pulse's qubit transition probability is simply $|v_{21}|^2$,  the difference between rectangular pulse and trapezoidal pulse can thus be estimated by the angle $\theta$. 

The parameter region where the Landau Zener contribution does not significantly affect the phase of the Rabi oscillation at the plateau can thus be bounded as:
\begin{align}\label{negligibleRampEq}
&\theta= 2\frac{g|\Delta|}{v}\ll 1
\end{align}
which gives
\begin{align}
    \Delta \ll \frac{v}{4g}  \approx \frac{\sqrt{250\text{MHz} \times 1\text{GHz}} }{20\text{MHz}}\approx 25\text{MHz}
\end{align}
where  $\Delta_1$  represents the absolute value of detuning at the plateau and  $\Delta_0$   represents the absolute value of initial qubit frequency. Given   $\Delta_0=600$MHz for the swap spectroscopy experiment, such condition is not satisfied for any data point been taken.
This explains why numerically rectangular pulse differs significantly from trapezoidal pulse in the   swap spectropy around TLS: even small amount of Landau Zener contribution can induce a non-negligible phase shift to the Rabi oscillation when the ramp time of the frequency tuning is long enough.

\section{Full Experimental 2D Swap-spectroscopy Data}\label{SuppData}

\begin{figure}[ht]
\begin{center}
\includegraphics[width=0.9\textwidth]{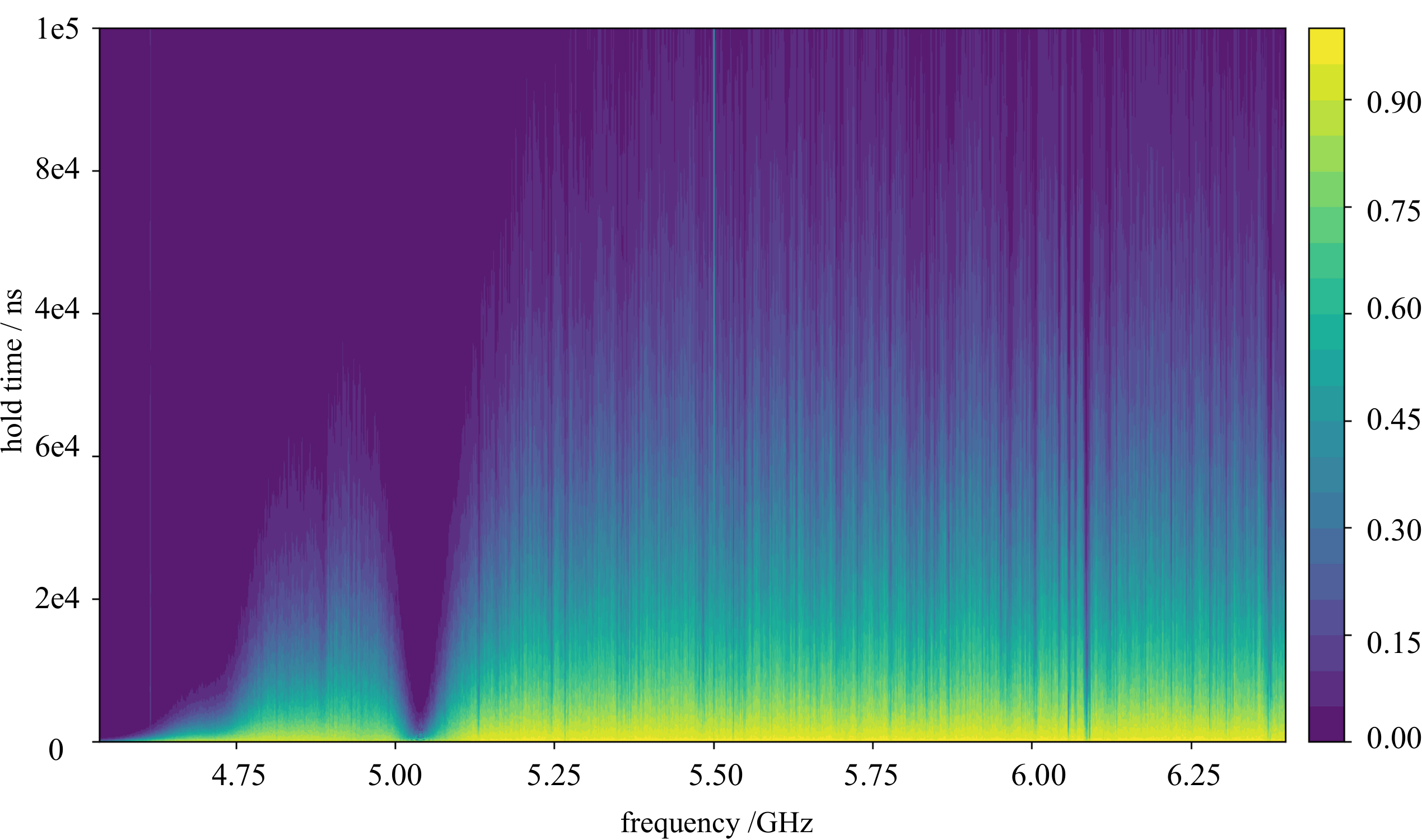}
\caption{Th complete TSSD  measured by the swap spectroscopy experiment. The readout resonator is around frequency of $4$GHz, and the qubit initial idel frequency is around $5.6$ GHz.
\label{fulldata}}
\end{center}
\end{figure}

The complete set of two-dimensional swap-spectroscopy scan is presented in Fig.~\ref{fulldata}. The highly coherent TLS that our characterization method is used to identify shows up as a blue dip near frequency 5.04GHz in the figure.  

\section{Moir\'{e} Effect }\label{MoireSupp}

\begin{figure}[H]
\begin{center}
\includegraphics[width=1.0\linewidth]{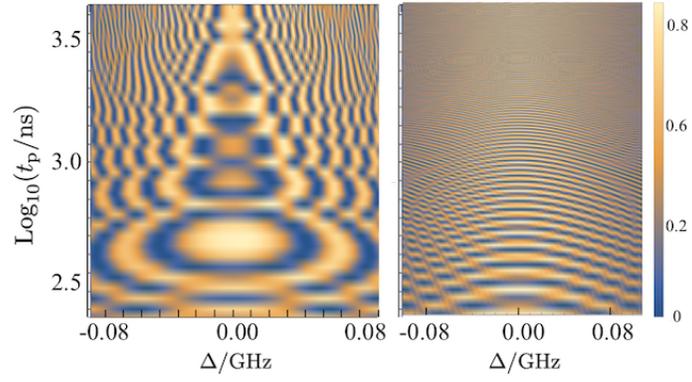}
\caption{Swap spectroscopy data predicted by fully coherent qubit-TLS coupling with $f_{\text{pl}}$ close to the \ftls with $\Delta = \fp -\ftls$ predicted by coherent qubit-TLS interaction model, $g=10$MHz, $\Gamma_2=10$MHz, $\tr = 6$ns.  Left plot: TSSD demonstrates additional circular oscillations with an increased period under a non-unform temporal sampling step. Right plot:  TSSD shows commonly observed Rabi-Chevron patterns under a uniform temporal sampling step.
\label{CompareMoireUniformFig}}
\end{center}
\end{figure}

Moir\'{e} effect occurs when two-periodic patterns overlay with each other to create a new pattern of a different periodicity. In our experiment, the first pattern is the original population oscillation between qubit and TLS in frequency and time. The second pattern   is the oscillation in time with a period increasing exponentially with time realized through non-uniform temporal sampling.  The effect of the non-uniform temporal sampling  on TSSD can then be understood by juxtaposing horizontal lines separated by an exponentially increasing spacing with the original Rabi-Chevron pattern of TSSD. As shown in our numerical simulation, see Fig.~\ref{CompareMoireUniformFig}), where adoping the experimentally relavant TLS parameter, the non-Markovian oscillations between qubit's and TLS's population is amplified by around three magnitudes~(shown as the  periodicity of the left plot is roughly a factor of three larger than that of  the right plot in log-scale at $\Delta = \fp - \ftls =0$.

\section{Experimental Non-idealities}\label{ExperimentNoiseSec}
\begin{figure}[H]
\begin{center}
\includegraphics[width=1\linewidth]{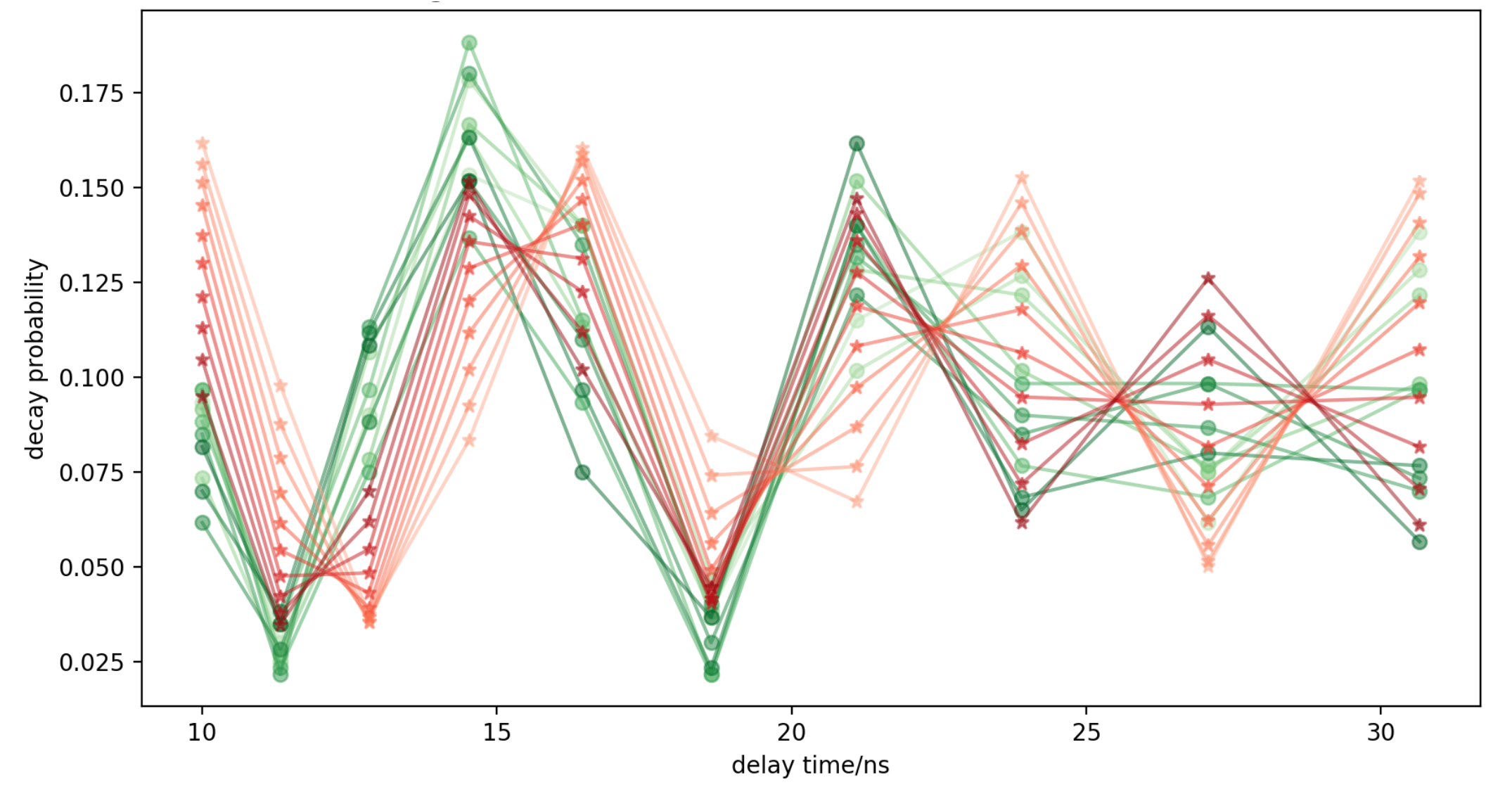}
\caption{ One dimensional data from TSSD and prediction from theory by fitting the 1D data in time at a fixed frequency: decay probability of qubit as a function of time for plateau frequency  $\fp \in [4.898$GHz,  $4.845$GHZ]. One dimensional data:  green dots with the opacity decreasing as frequency increases.  Prediction from fitted TLS model: red stars with the opacity decreasing as frequency increases. 
\label{NoisyFitting1DFigure}}
\end{center}
\end{figure}

Fitting a noisy experimental data is hard: I.   TSSD produced by drastically different  values of $\vec{p}_{\text{TLS}}$ can give rise to comparable cost due to measurement noise; II. the noisy data size is much larger than the number of model parameters resulting in underfitting.
For example, Fig.~\ref{NoisyFitting1DFigure} shows the fitting outcome of a TLS model for a given frequency, which provide a highly  unreliable prediction of data at other frequeny of TSSD. 
The fact that the model fitted to one the data taken at one frequency doesn't fit well to the data at a different frequency can be contributed to the experimental non-idealities including measurement errors and Purcell effect. Measurement error can be modeled by a biased blipflip error on the computational basis measurement outcoomes.   Qubit relaxation rate will increase when its frquency is close to the readout resonator at around 4 GHz. This is described by the Purcell effect which depends on the Jaynes-Cummings coupling between qubit and environmental bath $g_{JC}$,  the frequency difference $\Delta =| f_q-\omega_{r}|$, and the resontator life time $1/\kappa$, as $ \delta \Gamma_{q,1} = \kappa \frac{g_{JC}^2}{\Delta^2}.$ as~\cite{Schoelkopf2007}

\section{Operator Sum Description of   Two Qubit Gate}\label{SuppKraus}
With an explicit physical  TLS model,  in this section we derive the associated Kraus operator description for the realizable two-qubit gates in Google's   superconducting qubits~\cite{Barends2019}.  This facilitates the simulation of a realistic noisy circuit and the estimation of overall circuit fidelity of realistic quantum device. 

For simplificity, we consider the case where each qubit has one neighboring TLS. This assumption applies for our experimental setup since idel qubit frequency is always calibrated to be far far from known TLSs, but due to the temporal dynamics  during a two-qubit gate, the   qubit frequency might coincide with a neighboring qubit during its frequency tuning.  The joint system Hamiltonian for TLS and qubit takes the form:

\begin{align}
    \hat{H}(t)=\sum_{j=1}^2 \left[  \hat{H}_{TLS,j} + \hat{H}_{Q,j} + \hat{H}_{TLS,Q,j} \right]
\end{align}where each term is defined in Eq.~(\ref{EqTLSQUBITfirst})-(\ref{EqTLSQUBITlast}). 

\subsection{Two-Qubit Gate}
During a two-qubit gate, the system Hamiltonian of the qubit is described by the following Bose-Hubbard Hamiltonian:
\begin{align}\label{BSHamiltonian}
 \hat{H}_{BH} = \sum_{j=1,2}[f_q \hat{a}_j^\dagger \hat{a}_j - \frac{\eta_{BH}}{2} \hat{a}_j^\dagger \hat{a}_j(\hat{a}_j^\dagger \hat{a}_j-1)] +  g  (\hat{a}_1\hat{a}_2^\dagger + \hat{a}_1^\dagger \hat{a}_2), 
\end{align} where $\eta_{BH}$ represents the anharmoniticy of the nonlinear superconducting circuit oscillator. 
The unitary  induced by the  Bose-Hubbard Hamiltonian in Eq.~(\ref{BSHamiltonian}) in the single excitation subspace, i.e. the qubit subspace, $\mathcal{H}_q=$Span$\{\ket{0,0},\ket{0,1},\ket{1,0},\ket{1,1}\}$ is 
\begin{align}
U_2(g, t_2, f_q) & =\exp[-i \hat{H}_{BH} t_2] \\
&=\left(\begin{matrix}1 & 0& 0& 0\\
0 & e^{-i f_q t_2}\cos(g t_2) & -i \sin(g t_2) e^{i f_q t_2}& 0\\
0 & -i \sin(g t_2) e^{i  f_q t_2} &  e^{-i  f_q t_2}\cos(gt_2) & 0\\
0& 0& 0 & e^{-i 2 f_q t_2} w_{11}
\end{matrix}\right)
\end{align} where we have assumed that the two qubits share the exact same frequency during interaction.   The amplitude $| w_{11}|=1$ for perfect unitary where leakage induced by the $g$ coupling between $\ket{11}$ and $\ket{20}$ or $\ket{02}$ is zero. For a generic two-qubit gate time $t_2$, the two single excitation state amplitude depends on the coupling strength and the anharmonicity  $\eta_{BH}$ as
\begin{align}
    w_{11} =e^{\frac{it_2\eta_{BH}}{2}} \left( \cos[\frac{1}{2} t_2\sqrt{16 \lambda^2 + \eta_{BH}^2}] - \frac{i\eta_{BH} \sin[\frac{t_2}{2}\sqrt{16 \lambda^2 + \eta_{BH}^2}]}{\sqrt{16 \lambda^2 + \eta_{BH}^2}}\right).
\end{align} 
Consequently,  when $t_2$ is not chosen carefully, information leakage occurs. Henceforth, we study the parameter regimes where such leakage is zero by the careful choice of interaction frequency and gate time. We focus our attention on the error induced solely by the coupling between qubit and TLS as well as its environment during a two-qubit gate. Such quantum channel that maps the density operator of the two-qubit system from an initial state to the final state   by   Kraus operators $V_{k_1,k_2,k_3,k_4}$ as
\begin{align}
\rho(t_2) = \sum_{k_1, k_2,k_3,k_4 \in [0,1]} V_{k_1,k_2,k_3,k_4} \rho(0) V_{k_1,k_2,k_3,k_4}^\dagger.
\end{align} Here,  we use four bit subscript representation for the convenience of the  discussions to follow.
To simplify the analysis, we go into the interaction picture  defined by  the two-qubit gate control Hamiltonians excluding the qubit-TLS coupling.  In the interaction picture, the qubit $\alpha$'s $k$th Pauli operators becomes:
\begin{align}\label{InteractionPicQubitPaulis}
    \tilde{\sigma}_\alpha^k(t) = U_2(g, t , f_q)^\dagger \sigma_\alpha^k U_2(g, t, f_q).
\end{align}  
The TLS operators are transformed into:
\begin{align}\nonumber
    \tilde{\sigma}_{TLS_\alpha}^x(t) &= e^{- \Gamma_{TLS,\phi}t}\left( \exp[i\frac{\omega_{TLS,\alpha}(I_{\alpha, TLS}-\sigma_{TLS_\alpha}^z )}{2} ]\sigma_{TLS,\alpha}^x\exp[-i\frac{\omega_{TLS,\alpha}(I_{\alpha, TLS}-\sigma_{TLS_\alpha}^z )}{2} ]\right)\\\label{InteractionPicTLSPaulis1}& = e^{- \Gamma_{TLS, \phi}t}\left(\sigma_{TLS,\alpha}^x \cos (\omega_{TLS,\alpha}t) + \sigma_{TLS,\alpha}^y \sin(\omega_{TLS,\alpha} t)\right)\\\nonumber
        \tilde{\sigma}_{TLS_\alpha}^y(t) &= e^{- \Gamma_{TLS, \phi}t}\left(\exp[i\frac{\omega_{TLS,\alpha}(I_{\alpha, TLS}-\sigma_{TLS_\alpha}^z )}{2} ]\sigma_{TLS,\alpha}^y\exp[-i\frac{\omega_{TLS,\alpha}(I_{\alpha, TLS}-\sigma_{TLS_\alpha}^z )}{2} ]\right)\\\label{InteractionPicTLSPaulis2}& =e^{- \Gamma_{TLS, \phi}t}\left(\sigma_{TLS,\alpha}^y \cos (\omega_{TLS,\alpha}t) -\sigma_{TLS,\alpha}^x \sin(\omega_{TLS,\alpha} t)\right)
\end{align} where  the exponential decay factor comes from the fact that  TLSs are also coupled to their own Markovian bath. Such additional decay term in the exponent is proportional to TLS dephasing rate $\Gamma_{TLS,\phi}$. 
Notice that so far  we have not included qubit bath coupling effect, which is additive under the weak coupling limit.


In the interacting picture,  the TLS and qubit coupling takes  the following form:
\begin{align}
\tilde{H}_{Q-env}(t)= \sum_{\alpha = 1,2} \lambda_\alpha \frac{\tilde{\sigma}_\alpha^x(t) \tilde{\sigma}_{TLS,\alpha}^x(t) +   \tilde{\sigma}_\alpha^y(t)\tilde{\sigma}_{TLS,\alpha}^y(t) }{2}
\end{align}
where  the longitudinal coupling between TLS and qubit is averaged out through rotating wave approximation within the interaction picture.

We assume the coupling strength $\lambda_\alpha$ is weak enough: $\lambda_\alpha\cdot t \ll 1$, such that the  linear perturbation theory applies.  The   joint unitary transformation   can  be found approximately by Dyson series to the second order:
\begin{align}\label{EqDysonSeries}
U_{Q-env}(t, 0) = I - \frac{i}{\hbar}\int_0^t \tilde{H}_{Q-env}(t_1) d t_1 - \frac{1}{\hbar^2} \int_0^{t_1} d t_1 \int_0^{t_2} d t_2  \tilde{H}_{Q-env}(t_1)\tilde{H}_{Q-env}(t_2).
\end{align}
 Projecting this unitary onto  different orthogonal states of TLSs in turn gives all   independent Kraus operators.   
If we represent the index of the Kraus operators  with binary digit $k= (k_1, k_2, k_3, k_4)$  with each binary number $k_i \in \{0,1\}$ representing the state of one of the TLSs near each qubit,  each one of $2^4$ Kraus operator is labeled by the contracted TLS state as
\begin{align}
V_{k_1, k_2, k_3, k_4, 2} = \bra{k_1}_{TLS,1}\bra{k_2}_{TLS,2} U_{Q-env}(t, 0)\ket{k_3}_{TLS,1}\ket{k_4}_{TLS,2}.
\end{align}

Inserting Eq.~(\ref{interactionPicQubitOp})-(\ref{InteractionPicTLSPaulis2}) into the Dyson series of Eq.~(\ref{EqDysonSeries}) above, and tracing out the TLS, we obtain four non-zero components which correspond to four independent Kraus operators for the noisy two-qubit gate:
\begin{align}\nonumber
    E_{00}&=\bra{0}_{TLS,1}\bra{0}_{TLS,2} U_{Q-env}(t_2, 0)  \ket{0}_{TLS,1}\ket{0}_{TLS,2}\\\nonumber
    &=- \frac{1}{\hbar^2}\sum_{\alpha=1,2} \lambda_\alpha^2 \left[ \hat{X}_\alpha^\prime(t_2 ,\omega_{TLS,\alpha} , \Gamma_{TLS_\alpha,\phi}, f_q, g, \eta)\hat{X}_\alpha^\prime(t ,-\omega_{TLS,\alpha} , \Gamma_{TLS_\alpha,\phi}, f_q, g, \eta)\right.\\\nonumber
    &- i \hat{Y}_\alpha^\prime(t_2, \omega_{TLS,\alpha} , \Gamma_{TLS_\alpha,\phi}, f_q, g, \eta)\hat{X}_\alpha^\prime(t_2 ,-\omega_{TLS,\alpha} , \Gamma_{TLS_\alpha,\phi}, f_q, g, \eta) \\\nonumber
    &  + i \hat{X}_\alpha^\prime(t_2 ,\omega_{TLS,\alpha} , \Gamma_{TLS_\alpha,\phi}, f_q, g, \eta)\hat{Y}_\alpha^\prime(t_2, -\omega_{TLS,\alpha} , \Gamma_{TLS_\alpha,\phi}, f_q, g, \eta)\\\label{E00EqSupp}
    & \left. + \hat{Y}_\alpha^\prime(t_2, \omega_{TLS,\alpha} , \Gamma_{TLS_\alpha,\phi}, f_q, g, \eta)\hat{Y}_\alpha^\prime(t_2, -\omega_{TLS,\alpha} , \Gamma_{TLS_\alpha,\phi}, f_q, g, \eta)  \right],\\\nonumber
    E_{10}&=\bra{1}_{TLS,1}\bra{0}_{TLS,2} U_{Q-env}(t_2, 0)  \ket{1}_{TLS,1}\ket{0}_{TLS,2}\\\label{E10EqSupp}
    &= - \frac{i}{2\hbar} \lambda_1\left[\hat{X}_1^\prime(t_2, -\omega_{TLS,\alpha} , \Gamma_{TLS_\alpha,\phi}, f_q, g, \eta) + \hat{Y}_1^\prime(t_2, -\omega_{TLS,\alpha} , \Gamma_{TLS_\alpha,\phi}, f_q, g, \eta) \right] ,\\\nonumber
        E_{01}&=\bra{0}_{TLS,1}\bra{1}_{TLS,2} U_{Q-env}(t_2, 0)  \ket{0}_{TLS,1}\ket{1}_{TLS,2}\\\label{E01EqSupp}
         & =- \frac{i}{2\hbar} \lambda_2\left[\hat{X}_2^\prime(t_2, -\omega_{TLS,\alpha} , \Gamma_{TLS_\alpha,\phi}, f_q, g, \eta) + \hat{Y}_2^\prime(t_2, -\omega_{TLS,\alpha} , \Gamma_{TLS_\alpha,\phi}, f_q, g, \eta)  ,
    \right]
\end{align}
\begin{align}
      E_{11} &=\bra{1}_{TLS,1}\bra{1}_{TLS,2} U_{Q-env}(t_2, 0)  \ket{1}_{TLS,1}\ket{1}_{TLS,2}\\\nonumber 
      &= - \frac{1}{\hbar^2}\sum_{\alpha=1,2,\beta \neq \alpha} \lambda_\alpha g_\beta \left[ \hat{X}_\alpha^\prime(t ,-\omega_{TLS,\alpha} , \Gamma_{TLS_\alpha,\phi}, f_q, g, \eta)\hat{X}_\beta^\prime(t_2, -\omega_{TLS,\beta} , \Gamma_{TLS_\beta,\phi}, f_q, g, \eta)\right. \\\label{E11EqSupp}
      & \left.+ i \hat{Y}_\alpha^\prime(t_2, -\omega_{TLS,\alpha} , \Gamma_{TLS_\alpha,\phi}, f_q, g, \eta)\hat{X}_\alpha^\prime(t ,-\omega_{TLS,\alpha} , \Gamma_{TLS_\alpha,\phi}, f_q, g, \eta) \right],
\end{align}
where the interaction picture TLS-qubit Pauli operators after tracing out the incoherent  TLS are represented by $\hat{X}^\prime_\alpha$ and $\hat{Y}^\prime_\alpha$   as follows:

\resizebox{.7\linewidth}{!}{
  \begin{minipage}{\linewidth}
  
  \begin{align}
  & \hat{X}^\prime_\alpha(t, -\omega_{TLS,\alpha} , \Gamma_{TLS_\alpha,\phi}, f_q, g, \eta)\\\nonumber
    &=\left(\begin{matrix}
    0 & - i FST[g,\omega_{TLS,\alpha},\Gamma_{TLS_\alpha,\phi}, f_q, g,t] &    FCT[g,\omega_{TLS,\alpha},\Gamma_{TLS_\alpha,\phi}, f_q, g,t] & 0\\\nonumber
  i FST[g,\omega_{TLS,\alpha},\Gamma_{TLS_\alpha,\phi}, -f_q, g,t] &  0 & 0 &    i FCWT[g,\omega_{TLS,\alpha},\Gamma_{TLS_\alpha,\phi}, f_q, g,t, \eta]\\\nonumber
    FCT[g,\omega_{TLS,\alpha},\Gamma_{TLS_\alpha,\phi}, -f_q, g,t] & 0 & 0 & i FSWT[g,\omega_{TLS,\alpha},\Gamma_{TLS_\alpha,\phi}, f_q, g,t, \eta]\\\nonumber
    0 & FCWT[g,\omega_{TLS,\alpha},\Gamma_{TLS_\alpha,\phi},- f_q, g,t, - \eta] & -i FSWT[g,\omega_{TLS,\alpha},\Gamma_{TLS_\alpha,\phi}, -f_q, g,t, -\eta] & 0
    \end{matrix}\right)
\end{align}
\begin{align}
  & \hat{Y}^\prime_\alpha(t, -\omega_{TLS,\alpha} , \Gamma_{TLS_\alpha,\phi}, f_q, g, \eta)\\\nonumber
    &=\left(\begin{matrix}
    0 & -   FST[g,\omega_{TLS,\alpha},\Gamma_{TLS_\alpha,\phi}, f_q, g,t] &    -i FCT[g,\omega_{TLS,\alpha},\Gamma_{TLS_\alpha,\phi}, f_q, g,t] & 0\\\nonumber
-FST[g,\omega_{TLS,\alpha},\Gamma_{TLS_\alpha,\phi}, -f_q, g,t] &  0 & 0 &    -i FCWT[g,\omega_{TLS,\alpha},\Gamma_{TLS_\alpha,\phi}, f_q, g,t, \eta]\\\nonumber
  i  FCT[g,\omega_{TLS,\alpha},\Gamma_{TLS_\alpha,\phi}, -f_q, g,t] & 0 & 0 & i FSWT[g,\omega_{TLS,\alpha},\Gamma_{TLS_\alpha,\phi}, f_q, g,t, \eta]\\\nonumber
    0 & i FCWT[g,\omega_{TLS,\alpha},\Gamma_{TLS_\alpha,\phi},- f_q, g,t, -\eta] &   FSWT[g,\omega_{TLS,\alpha},\Gamma_{TLS_\alpha,\phi}, -f_q, g,t, -\eta] & 0
    \end{matrix}\right)
\end{align}
\begin{align}
  & \hat{X}^\prime_2(t, -\omega_{TLS,\alpha} , \Gamma_{TLS_\alpha,\phi}, f_q, g, \eta)\\\nonumber
    &=\left(\begin{matrix}
    0 &  FCT[g,\omega_{TLS,\alpha},\Gamma_{TLS_\alpha,\phi}, f_q, g,t] &    -i FST[g,\omega_{TLS,\alpha},\Gamma_{TLS_\alpha,\phi}, f_q, g,t] & 0\\\nonumber
 FCT[g,\omega_{TLS,\alpha},\Gamma_{TLS_\alpha,\phi}, -f_q, g,t] &  0 & 0 &    i FSWT[g,\omega_{TLS,\alpha},\Gamma_{TLS_\alpha,\phi}, f_q, g,t, \eta]\\
    FST[g,\omega_{TLS,\alpha},\Gamma_{TLS_\alpha,\phi}, -f_q, g,t] & 0 & 0 &   FCWT[g,\omega_{TLS,\alpha},\Gamma_{TLS_\alpha,\phi}, f_q, g,t, \eta]\\
    0 & -i FSWT[g,\omega_{TLS,\alpha},\Gamma_{TLS_\alpha,\phi},- f_q, g,t, - \eta] & -i FCWT[g,\omega_{TLS,\alpha},\Gamma_{TLS_\alpha,\phi}, -f_q, g,t, -\eta] & 0
    \end{matrix}\right)
\end{align}
\begin{align}
  & \hat{Y}^\prime_2(t, -\omega_{TLS,\alpha} , \Gamma_{TLS_\alpha,\phi}, f_q, g, \eta)\\
    &=\left(\begin{matrix}
    0 &  -i FCT[g,\omega_{TLS,\alpha},\Gamma_{TLS_\alpha,\phi}, f_q, g,t] &    -  FST[g,\omega_{TLS,\alpha},\Gamma_{TLS_\alpha,\phi}, f_q, g,t] & 0\\
 i FCT[g,\omega_{TLS,\alpha},\Gamma_{TLS_\alpha,\phi}, -f_q, g,t] &  0 & 0 &     FSWT[g,\omega_{TLS,\alpha},\Gamma_{TLS_\alpha,\phi}, f_q, g,t, \eta]\\
  - FST[g,\omega_{TLS,\alpha},\Gamma_{TLS_\alpha,\phi}, -f_q, g,t] & 0 & 0 &   - i FCWT[g,\omega_{TLS,\alpha},\Gamma_{TLS_\alpha,\phi}, f_q, g,t, \eta]\\
    0 &  FSWT[g,\omega_{TLS,\alpha},\Gamma_{TLS_\alpha,\phi},- f_q, g,t, - \eta] & i FCWT[g,\omega_{TLS,\alpha},\Gamma_{TLS_\alpha,\phi}, -f_q, g,t, -\eta] & 0
    \end{matrix}\right)
\end{align}  \end{minipage}
}

with each element defined as
\begin{align}
 &   FST[g,\omega_{TLS,\alpha},\Gamma_{TLS_\alpha,\phi}, f_q, g,t]=\int_0^t \xi_{x,x,\alpha}(\tau) e^{-i \tau \omega_{TLS,\alpha}  }e^{-i \tau \omega_{q}  }\sin[g \tau] d\tau\\
  &   FCT[g,\omega_{TLS,\alpha},\Gamma_{TLS_\alpha,\phi}, f_q, g,t]=\int_0^t \xi_{x,x,\alpha}(\tau) e^{-i \tau \omega_{TLS,\alpha}  }e^{-i \tau \omega_{q}  }\cos[g \tau] d\tau\\
  &   FSWT[g,\omega_{TLS,\alpha},\Gamma_{TLS_\alpha,\phi}, f_q, g,t]=\int_0^t \xi_{x,x,\alpha}(\tau) e^{-i \tau \omega_{TLS,\alpha}  }e^{-i \tau \omega_{q}  }\sin[g \tau]w_{11}(\tau) d\tau\\
  &   FCWT[g,\omega_{TLS,\alpha},\Gamma_{TLS_\alpha,\phi}, f_q, g,t]=\int_0^t \xi_{x,x,\alpha}(\tau) e^{-i \tau \omega_{TLS,\alpha}  }e^{-i \tau \omega_{q}  }\cos[g \tau]w_{11}(\tau) d\tau,
 \end{align} where $\xi_{x,x,\alpha}(\tau)=\langle \sigma_{TLS,\alpha}^x(\tau)\sigma_{TLS,\alpha}^x(0) \rangle$ is the correlator~(unsymmetrized) given by the response function of TLSs. We have chosen an exponential decay function for this correlator $\xi_{x,x,\alpha}(\tau)=\exp[-\tau \Gamma_{TLS,\phi} ]$ in Eq.~(\ref{InteractionPicTLSPaulis1}) and Eq.~(\ref{InteractionPicTLSPaulis2}) for our case where each TLS is coupled to a Markovian environment.  Such assumptions is not essential to our analysis since we can replace the correlator by any form in principle, including that induced by the coupling to a non-Markovian environment.

  We list  the amplitudes for each two-qubit Pauli operator for all non-zero Kraus operators in Table.~\ref{Table2Qkraus}, where we use short-handed notations:
\begin{align}
    & FST_\alpha :=FST[g,-\omega_{TLS,\alpha},\Gamma_{TLS_\alpha,\phi}, f_q, g,t]\\
    & FCT_\alpha :=FCT[g,-\omega_{TLS,\alpha},\Gamma_{TLS_\alpha,\phi}, f_q, g,t]\\
    &FSWT_\alpha :=FSWT[g,-\omega_{TLS,\alpha},\Gamma_{TLS_\alpha,\phi}, f_q, g,t,\eta]\\
    &FCWT_\alpha :=FCWT[g,-\omega_{TLS,\alpha},\Gamma_{TLS_\alpha,\phi}, f_q, g,t,\eta],\\
    &\beta = FCT_\alpha FST_\alpha^*.
\end{align}

\begin{table} [ht] 
\centering  
\scalebox{0.65}{%
\begin{tabular}{|c|c|c|c|c|c|c|c|}
\cline{1-8} 
 &  \multicolumn{1}{ |c| }{ $\sigma_1^z\sigma_2^z$} & \multicolumn{1}{ |c| }{$\sigma_1^x \sigma_2^x + \sigma_1^y\sigma_2^y$ }  & \multicolumn{1}{ |c| }{$\sigma_1^x \sigma_2^x - \sigma_1^y\sigma_2^y$ } & \multicolumn{1}{ |c| }{$\sigma_1^x \sigma_2^y - \sigma_1^y\sigma_2^x$ }& \multicolumn{1}{ |c| }{$\sigma_1^x \sigma_2^y + \sigma_1^y\sigma_2^x$ }& \multicolumn{1}{ |c| }{$\sigma_1^z \sigma_2^x + i  \sigma_1^z\sigma_2^y$ } & \multicolumn{1}{ |c| }{$\sigma_1^x \sigma_2^z + i  \sigma_1^y\sigma_2^z$ }\\ \cline{1-8}
$E_{00}$ & $\lambda_{1}^2\vert FST_1\vert^2+\lambda_2^2\vert FCT_2\vert^2 $ 
     & $i\frac{1 }{2}(\lambda_1^2 Re[\beta] + \lambda_2^2 Im[\beta]) $& &$i\frac{ 1}{2}( \lambda_1^2 Im[\beta] + \lambda_2^2 Re[\beta] )$& &&  \\ \cline{1-8}
$E_{01}$   &    
    &     &&&& $\frac{1 }{4}\lambda_2(FCT_2-FCWT_2)$ &$-i\frac{1 }{4}\lambda_2(FST_2+FSWT_2)$ \\ \cline{1-8}
    $E_{10}$   &    
    &     &&&& $\frac{1 }{4}\lambda_1(-i FST-FSWT)$ & $\frac{1 }{4}\lambda_1(FCT_1-FCWT_1)$ \\ \cline{1-8}
    $E_{11}$   &    
    &     & $\frac{1}{4}(FST_1 \cdot FSWT_2 - FCWT_1 \cdot FCT_2)$ & & $ i\frac{1}{4}(FST_1 \cdot FSWT_2 - FCWT_1 \cdot FCT_2)$& &  \\ \cline{1-8}
\end{tabular}}
\caption{Amplitudes for different Kraus Operators.}\label{Table2Qkraus}
\end{table}

\section{Fidelity Estimate}\label{FidelityEstimate}
In this section, we introduce basic measures for the quality of the quantum channel: the  average gate fidelity and the rescaled unitarity first defined in~\cite{wallman2015estimating}. The former measures the total amount of gate error in average case, and the latter measures the contribution to this total error from purely decoherent effects. It is proven in \cite{wallman2015estimating} that the rescaled unitarity upper bounds the average gate fidelity, which bound is tight when the unitary error is exactly zero.

Once we know the specific form of the noisy channel Kraus operators, the average fidelity of each quantum gate $U$ can be evaluated through the simple relation between entanglement fidelity $F_e$ and average fidelity $F_{ave}$ as:
\begin{align}\label{averageFidelity}
    F_{ave}=\frac{dF_e +1}{d+1}.
\end{align}
Let us represent the ideal  quantum gate on $n$-qubit by a $d\times d$ dimensional unitary $U$ with $d=2^n$, and the full set of  $n$-qubit  Pauli operators  represented by $G_j$ with $j \in[d^2]$,   the entanglement fidelity can be expressed by the sum of   Kraus operators as:
\begin{align}\label{EntanglementFidelity}
    F_e =\frac{\sum_j \text{Tr}\left[U G_j^\dagger U^\dagger \sum_k V(k) G_j V(k)^\dagger \right]}{d^2}=\frac{\sum_k \left\vert\text{Tr}\left[UV(k)^\dagger\right]\right\vert^2}{d^2}
\end{align}   
 
We evaluate the average gate fidelity of a two-qubit gate $\sqrt{\text{ISWAP}}$ defined in Eq.~(\ref{SqrtISWAP}) under the realistic noise channel under TLS-qubit coupling described by Eq.~(\ref{E00EqSupp})-(\ref{E11EqSupp}).
\begin{align}\label{SqrtISWAP}
  \sqrt{\text{ISWAP}}=  \left(
\begin{array}{cccc}
 1 & 0 & 0 & 0 \\
 0 & \frac{1}{\sqrt{2}} & \frac{i}{\sqrt{2}} & 0 \\
 0 & \frac{i}{\sqrt{2}} & \frac{1}{\sqrt{2}} & 0 \\
 0 & 0 & 0 & 1 \\
\end{array}
\right)
\end{align}
We show in Fig.~5 of the main text that  the average fidelity of two-qubit gate is minimal near the dressed interaction frequency  $f_q \pm g$. Such gate frequency dependence is a distinct nature from non-Markovian type of errors.

The same Kraus operator representation also help us to directly evaluate unitarity of a quantum channel $\epsilon$ defined by:
\begin{align}\label{unitarityEQ}
    u(\epsilon) =\frac{d-1}{d}\int d \psi \epsilon^\prime(\psi)^\dagger \epsilon^\prime(\psi)
\end{align}
with $\epsilon^\prime(\rho)= \epsilon(\rho) - \text{Tr}[\epsilon(\rho)/\sqrt{d}] \mathbb{I} $. Since similar to the average gate fidelity, this quantity is also a second order polynomial in the gate and its complex conjugate, it can be evaluated using unitary 2-design, where the average over Harr measure in Eq.~\ref{unitarityEQ} can be replace by the average over Clifford group element. Then the rescaled non-unitary is defined as 
\begin{align}
    u^\prime (\epsilon) =\left( 1- \sqrt{ u(\epsilon)}\right)\frac{d-1}{d},
\end{align} which provides the measure for the incoherent contribution to the average gate error and equals the average gate error when the noise channel is purely decoherence.





\section{Evolutionary Algorithm}\label{SuppEA}
In this subsection, we described the detailed implementation of EA in our experiments. The detailed algorithm is defined in \ref{EAALGORIGHTM}.

We use neural network as a function approximator to map a given TSSD data set $D_{TSSD}$ to the TLS model parameters $ \vec{p}_{TLS}$ which reproduce the observed data through our physical model in Eq.~(5) of the main text. The neural network takes   $D_{TSSD}$ as input and output $ \vec{p}_{TLS}$ through forward propagation defined by the iterative updates between hidden variables, i.e.,   the  $l+1$th layer hidden variable $Y_{l+1}$ depends on the previous layer as:
\begin{align}
    Y_{l+1} = \sigma_{non} \circ (W^l Y_l + \mathbf{b}^l)
\end{align}
where $\sigma_{non}\circ()$ represents the point-wise nonlinear function applied to each one of the vector element $v_i$. The EA updates defined in \ref{EAALGORIGHTM} iteratively changes the neural network weights $\{W^l\}$ and biases $\{\mathbf{b}^l\}$ such that given $Y_0=D_{TSSD}$ we have $Y_n=\vec{p}_{TLS}$ for an $n$ layer neural network. 



More specifically, we choose a three layer neural networks with width $10, 30, 12$ to accept a two dimensional data described by a $40\times 20$ matrix representing the decay probability for 40 different plateau frequencies $f_{\text{p}}$ and 20 different hold times \tp as input, to output the TLS parameters that minimizes the L2 norm between predicted two-dimensional data and the measured one. Three sets of such data shown as the first row of    Fig.3 of main text is used.  

We train the neural network with EA, where we choose the hyper-parameters defined in the algorithm \ref{EAALGORIGHTM} as follows.
\begin{align}
  & \mathbf{a}_{W}=\mathbf{a}_{B} =0.0031 \times \vec{1},\\
  & \sigma_W =\sigma_B =0.01,\\
  & N=100,\\
  &  M=50,\\
  & D=3\end{align} where we use $ \vec{1}$ to represent a vector of the same size as $ \mathbf{a}_{W}$ with each entry equal to $1$.
    
  \begin{algorithm}\captionsetup{labelfont={sc,bf}, labelsep=newline}
  \caption{Evolutionary algorithm for learning TLS parameters.\label{EAALGORIGHTM}} 
\begin{algorithmic} 
  \STATE{\textbf{Input}: Dimension of the neural network specified by the vector $\mathbf{V}=\{v_1, \ldots, v_n\}$, whose length specifies the number of layers, whose element specifies the number of neurons per layers.

 A set of $D$ two-dimensional data: $\{ D_{TSSD}^1, \ldots, D_{TSSD}^D\}$ of different range of \fp and \tp.
 
 Initial value for the weight matrix $W^l_0 $ and bias vector $\mathbf{b}^l_0 $ for the $l$th hidden layer.
 
Total optimization steps $N$,   the mean and variances of the perturbation for neural network's weights~(denoted by subscript $W$) and biases~(denoted by subscript $B$): $\mu_{W}=0, \sigma_{W},\mu_B=0, \sigma_B$, learning rates  for neural network's weights~(denoted by subscript $W$) and biases~(denoted by subscript $B$): $ \{ \mathbf{a}_W,\mathbf{a}_B\}$, evolution batch size $b$. Reward function $f(D_{TSSD}, \vec{p}_{TLS})$ for a given set of TSSD data $D_{TSSD}$ and a set TLS parameters $\vec{p}_{TLS} =\{\lambda, \ftls, \tp, \Gamma_{TLS, \phi}\}$\\
 \textbf{Output}: An estimate of $\vec{p}_{TLS}$ that minimize the cost function $f(D_{TSSD}, \vec{p}_{TLS})$.}
 
\STATE{Iterate through different two-dimensional TSSD data}:
\FOR{$h$ in $D$}
\STATE{For each set of data, optimize neural network with following iterations:}
\FOR{$k$ in $N$}
         
    \FOR{ each   $j$ in $b$}
    \STATE{Update the neural networks by adding  perturbations to the weight and bias of  each $l$ of $n$ layers by: }
    \FOR{ each   $l$ in $n$ layers}
 \STATE{Sample perturbations: $ \xi_{W^l}^j $ from the normal distribution $\mathcal{N}(\mu_W, \sigma_W)$. }
  \STATE{Sample perturbations: $ \xi_{\mathbf{b}^l}^j $ from the normal distribution $\mathcal{N}(\mu_B, \sigma_B)$. }
  \STATE{\begin{align}
      &W_{k, j}^l=\bar{W}_{k-1}^l +\xi_{W^l}^j\\
         &\mathbf{b}_{k, j}^l=\bar{\mathbf{b}}_{k-1}^l +\xi_{\mathbf{b}^l}^j
  \end{align}}  
  \ENDFOR 
  \STATE{Obtain the TLS parameters predicted by the current neural network from data through forward propagation: $\vec{p}_{TLS}^j$}
    \ENDFOR 
   
    \STATE{Obtain the neural network weights and biases through weighted average of perturbations:}
    \FOR{ each   $l$ in $n$ layers}
  \STATE{ \begin{align}
        &\bar{W}_{k }^l=\bar{W}_{k-1}^l+ \mathbf{a}_W \cdot\left(\frac{1}{b}\sum_{j=1}^b f(D_{TSSD}^h, \vec{p}_{TLS}^j) \xi_{W^l}^j\right)\\
      &\bar{\mathbf{b}}_{k }^l=\bar{\mathbf{b}}_{k-1}^l+ \mathbf{a}_B \cdot\left(\frac{1}{b}\sum_{j=1}^b  f(D_{TSSD}^h, \vec{p}_{TLS}^j) \xi_{\mathbf{b}^l}^j\right)\\
    \end{align} }
      \ENDFOR
      \ENDFOR
\ENDFOR  
\end{algorithmic}
\end{algorithm}
\end{appendix}

\end{document}